\documentclass[a4paper,12pt]{article}
\usepackage[utf8]{inputenc}

\usepackage{mathrsfs}
\usepackage{a4wide,latexsym,graphicx}
\usepackage{amsmath,amssymb}
\usepackage{hyperref}
\usepackage[numbers,sort&compress]{natbib}

\begin{document} 
\parskip=3pt plus 1pt

\begin{titlepage}
\hspace*{\fill}IFIC/21-01
\vskip 2cm	

\begin{center} 
\begin{LARGE} \bf\boldmath 
SU(3) analysis of four-quark operators:\\[5pt] $K\to\pi\pi$ and vacuum matrix elements
\end{LARGE}
\\[40pt] 
A. Pich${}^{1)}$ and A. Rodr\'iguez-S\'anchez${}^{2)}$
 
\vspace{1cm}
${}^{1)}$ Departament de F\'{\i}sica Te\`orica, IFIC, CSIC --- 
Universitat de Val\`encia \\ 
Edifici d'Instituts de Paterna, Apt. Correus 22085, E-46071 
Val\`encia, Spain \\[10pt]

${}^{2)}$ Université Paris-Saclay, CNRS/IN2P3, IJCLab, 91405 Orsay, France \\[10pt]  

\end{center} 
 
\vfill 

\abstract{Hadronic matrix elements of local four-quark operators play a central role in non-leptonic kaon decays, while vacuum matrix elements involving the same kind of operators appear in inclusive dispersion relations, such as
those relevant in $\tau$-decay analyses. Using an $SU(3)_L\otimes SU(3)_R$ decomposition of the operators, we derive generic relations between these matrix elements, extending well-known results that link observables in the two different sectors.
Two relevant phenomenological applications are presented. 
First, we determine the electroweak-penguin contribution to the kaon CP-violating ratio $\varepsilon'/\varepsilon$, using the measured hadronic spectral functions in $\tau$ decay. Second, we fit our $SU(3)$ dynamical parameters to the most recent lattice data on $K\to\pi\pi$ matrix elements. The comparison of this numerical fit with results from previous analytical approaches provides an interesting anatomy of the $\Delta I = \frac{1}{2}$ enhancement, confirming old suggestions about its underlying dynamical origin.}

\vfill
 
\end{titlepage} 

\section{Introduction}
\label{sec:intro}
 
Local operators with dimension larger than four, such as four-quark operators, play a key role in quantitatively understanding 
the low-energy dynamics of renormalizable theories. When working with a quantum field theory involving widely-separated scales, such as the Standard Model (SM), the logarithms of large scale ratios induce higher-order corrections that slow-down, if not directly spoil, the standard perturbative series. The use of short-distance techniques like the Operator Product Expansion (OPE) \cite{Wilson:1969zs} to separate scales and Renormalization Group Equations (RGE) to resum those logarithmic corrections becomes then a must \cite{Buchalla:1995vs,Pich:1998xt,Manohar:2018aog,Neubert:2019mrz}. When these techniques are applied, the resulting Effective Field Theory (EFT) contains a series
of low-energy operators, whose quantitative role in a given observable is, in general, inversely proportional to their dimensions.

At low energies, the short-distance logarithmic resummation is not enough. Owing to confinement and the associated growing of the strong coupling, the low-energy theory cannot be formulated in terms of approximately-free quarks and gluons; the relevant degrees of freedom are, instead, hadrons.
In practice, one runs perturbatively the EFT to energies as small as possible, so that all large short-distance logarithms can be reabsorbed into the computed Wilson coefficients, but the hadronic matrix elements of their associated operators must still be determined
with non-perturbative methods.

At very low energies, the only observed hadrons are pions, kaons and eta bosons. Due to their flavour structure, non-leptonic kaon decays cannot occur through strong or electromagnetic interactions; one needs to trace back their origin to the only source of flavour-breaking in the SM, the $W$ boson, whose imprint in the effective low-energy Lagrangian appears through dimension-six four-quark operators. The non-perturbative calculation of the corresponding hadronic matrix elements is a formidable task and current theoretical uncertainties for the associated observables are unfortunately large
\cite{Cirigliano:2011ny}. Improved lattice computations, {\it e.g.}, see \cite{Abbott:2020hxn}, may change the situation in the future.

A more precise knowledge arises in inclusive semileptonic processes
involving three light quark flavours, such as hadronic tau decays or electron-positron annihilation into hadrons \cite{Pich:2020gzz}. 
Although they have a quite different nature, the former being a weak-interaction transition and the latter an electromagnetic one, their
associated hadronic distributions can be studied with the same theoretical formalism, since rigorous dispersion relations \cite{Kallen:1952zz,Lehmann:1954xi} connect them with two-point correlation functions of quark currents, leading to very precise predictions \cite{Braaten:1991qm}. It is precisely in the OPE of these two-point correlation functions \cite{Shifman:1978bx} where the four-quark operators appear. In the same way that local quark operators can give non-zero matrix elements in transitions among hadrons, they can also acquire non-vanishing expectation values in the non-perturbative QCD vacuum. A well-known example is the $\bar q q$ condensate that plays a key role in the dynamical breaking of chiral symmetry. Unlike in non-leptonic kaon decays, the numerical role of four-quark operators is very small in the $\tau$ decay width because they enter suppressed by six powers of the tau mass. Nevertheless, with the achieved experimental accuracy, it is possible to extract significant dynamical information on some operators from the current $\tau$ data samples.

Non-trivial relations among matrix elements involving different four-quark operators can be derived, using their known symmetry transformations together with our knowledge of strong interactions at low energies. Many of these relations have been exploited in the past, but they appear somehow scattered in the literature~\cite{Donoghue:1982cq,Bijnens:1984ec,Pich:1985ab,Bernard:1985wf,Guberina:1985md,Pich:1985st,Kambor:1989tz,Pich:1990mw,Kambor:1991ah,Kambor:1992he,Pich:1995qp,Knecht:1998nn,Donoghue:1999ku,Bijnens:2001ps,Knecht:2001bc,Cirigliano:2001qw,Cirigliano:2002jy,Cata:2003mn,Hambye:2003cy}. In the following, we aim to provide a self-contained derivation, based only on symmetry considerations and EFT, 
and apply them to the phenomenology of non-leptonic kaon decays. As an important application, we will determine the electromagnetic-penguin contribution to $\varepsilon'/\varepsilon$, using the measured hadronic spectral functions in $\tau$ decay. Our determination will be compared with the updated values obtained combining Chiral Perturbation Theory ($\chi$PT) and large-$N_C$ techniques \cite{Gisbert:2017vvj,Cirigliano:2019cpi},
and with the most recent lattice results \cite{Abbott:2020hxn}.

We will also present a global fit to the available lattice data on $K\to\pi\pi$ matrix elements \cite{Abbott:2020hxn,Blum:2015ywa,Blum:2012uk}, in terms of a complete set of independent dynamical parameters with well-defined $SU(3)_L\otimes SU(3)_R$ transformation properties, at next-to-leading order (NLO) in $\alpha_s$ (short-distance logarithms) and $\chi$PT. The comparison of this numerical fit with previous analytical results makes possible to achieve a quantitative assessment of the different approximations adopted in those approaches. This provides an interesting anatomy of the $\Delta I=\frac{1}{2}$ enhancement, confirming old suggestions about its underlying dynamical origin.

The paper is organized as follows. Section~\ref{sec:4quark_EFT} focuses on the derivation of symmetry relations, making use of effective Lagrangians. The formalism is applied to strangeness-changing transitions in Section~\ref{sec:DStransitions}, which recovers the usual notation employed in $\chi$PT~\cite{Cirigliano:2011ny}. In Section~\ref{sec:condensates}, we apply the same tools to analyze the four-quark vacuum condensates appearing in the correlation functions of the QCD currents. This provides the wanted connection between the two sectors, making possible to determine with $\tau$ data a non-perturbative dynamical parameter characterizing the electroweak-penguin operator $Q_8$. This determination is presented in Section~\ref{sec:pheno}, after introducing all necessary dispersive tools.
A phenomenological analysis of $K\to\pi\pi$ matrix elements is presented in Section~\ref{sec:latt}, which contains the implications of our dispersive result for $\varepsilon'/\varepsilon$ and the numerical fit to the most recent lattice data.
A detailed discussion of our current understanding of the $\Delta I=\frac{1}{2}$ rule is given there, based on the fitted results and
the previous analytical knowledge. As a final consistency check, we also provide a precise determination of the pion decay constant, combining the parameters fitted to the lattice data with the measured inclusive distribution of the final hadrons in $\tau$ decay.
The main results of our paper are finally summarized in Section \ref{sec:conclusions}.

\section{Low-energy realization of four-quark operators}
\label{sec:4quark_EFT}

The massless QCD Lagrangian with three quark flavours, 
\begin{equation}\label{eq:LQCD0}
{\mathcal{L}}_{\mathrm{QCD}}^0 = -\frac{1}{4}\, G^a_{\mu\nu} G^{\mu\nu}_a
 + i \,\bar q_L^{\phantom{\dagger}} \gamma^\mu D_\mu q_L^{\phantom{\dagger}}  + i \,\bar q_R^{\phantom{\dagger}} \gamma^\mu D_\mu q_R^{\phantom{\dagger}}
\end{equation}
with $q^T = (u,d,s)$,
is invariant under $(L,R) \in SU(3)_{L}\otimes SU(3)_{R}$ global transformations in the flavour space: $q'_{L,i} = L_{i}^{\ j} q_{L,j}$, $q'_{R,i} = R_{i}^{\ j} q_{R,j}$, where $q_L = \frac{1}{2}\, (1-\gamma_5)\, q$ and $q_R = \frac{1}{2}\, (1+\gamma_5)\, q$ denote the left and right quark chiralities. This chiral symmetry is however not seen in the hadronic spectrum, which is only invariant under $SU(3)_V$ transformations with $L=R$. Thus, chiral symmetry is dynamically broken by the QCD vacuum, giving rise to eight $0^-$ massless Goldstone bosons that can be identified with the lightest pseudoscalar octet ($\pi$, $K$, $\eta$).

Together with parity ($P$) and charge-conjugation ($C$) invariance, chiral symmetry enforces very strong constraints on the low-energy dynamics of these (pseudo)Goldstone bosons that can be most easily analyzed with an effective Lagrangian expanded in powers of derivatives~\cite{Pich:2018ltt}. A convenient parametrization of the Goldstone fields is provided by the unitary matrix 
$U[\phi_{i}]=e^{i\lambda_{i}\phi_{i}/F}$, transforming as $U\rightarrow R\,  U L^{\dagger}$ under chiral rotations. At leading order (LO) in the derivative expansion, the effective Goldstone Lagrangian contains only two terms \cite{Gasser:1984gg}:
\begin{equation}\label{eq:LChPT}
\mathcal{L}_{\mathrm{eff}}\, =\, \frac{F^2}{4}\, \mathrm{Tr}( D_\mu U^\dagger D^\mu U + U^\dagger\chi + \chi^\dagger U)\, + \mathcal{O}(p^4)\, .
\end{equation}
The covariant derivative $D^\mu U =\partial^\mu U - i r^\mu U + i U\ell^\mu$ includes auxiliary external left ($\ell^\mu$) and right ($r^\mu$) matrix-valued vector sources coupled to the quarks, which allow us to easily derive the low-energy realization of the QCD currents~\cite{Pich:2018ltt}. The second term incorporates the couplings to external scalar ($s$) and pseudoscalar ($p$) sources through $\chi = 2 B_0\, (s + i p)$. Taking $p=0$ and $s=\mathcal{M}=\mathrm{diag}(m_u,m_d,m_s)$, this term implements the explicit breaking of chiral symmetry induced by the non-zero quark masses, generating the physical masses of the eight pseudoscalar bosons. The LO effective Lagrangian $\mathcal{L}_{\mathrm{eff}}$ completely determines the $\mathcal{O}(p^2)$ contributions to the Goldstone masses and scattering amplitudes, in terms of the quark masses, and the two low-energy couplings (LECs) $F$ and $B_0$, which are related to the pion decay constant and the $\bar q q$ vacuum condensate \cite{Pich:1995bw}. 

The underlying QCD Lagrangian, including the external sources $\ell^\mu$, $r^\mu$, $s$ and $p$, and its low-energy Chiral Perturbation Theory ($\chi$PT) \cite{Gasser:1984gg,Weinberg:1978kz} realization $\mathcal{L}_{\mathrm{eff}}$ are connected through the path integral expression
\begin{align}\nonumber
\exp{\{i Z[\ell^\mu, r^\mu, s, p]\}} \, &= \,  \int  \mathcal{D} q \,\mathcal{D} \bar q
\,\mathcal{D} G_\mu \,
\exp{\left\{i \int d^4x\, \mathcal{L}_{\mathrm{QCD}}\right\}}  
\, \\&=\,
\int  \mathcal{D} U \,
\exp{\left\{i \int d^4x\, \mathcal{L}_{\mathrm{eff}}\right\}} \, .\label{eq:generatingfunctional}
\end{align}
By taking functional derivatives with respect to the appropriate external sources in both terms of the equality, one finds the explicit low-energy expressions of the QCD quark currents. This 
dictionary will be exploited below to derive some useful relations among four-quark operators. Many of those symmetry relations are well-known, although quite often they are presented without a crystal-clear derivation or resorting to soft-pion methods. The next subsections compile them together, using a much simpler approach purely based on symmetry arguments.

\subsection{Chiral symmetry decomposition}
\label{subsec:chiral}

At low energies below the electroweak scale $v$, the renormalizable SM gives rise to an effective short-distance Lagrangian that contains dimension-six four-fermion operators. They originate from integrating out the heavy degrees of freedom ($t$, $H$, $Z$, $W^\pm$, $b$, $c$), which is needed in order to resum the large perturbative logarithms generated by the sizeable ratios of mass scales~\cite{Buchalla:1995vs,Pich:1998xt,Manohar:2018aog,Neubert:2019mrz}. The phenomenological effects of these `irrelevant' electroweak operators are suppressed by a factor $E^2/v^2\sim G_F E^2$, where $E$ is the energy scale of the process. They can then be treated as small perturbations to the QCD Lagrangian, in the sense that it is usually enough to analyze their implications to LO in $G_F$.

Let us then consider the extended QCD Lagrangian
\begin{equation}\label{eq:sourllll}
\mathcal{L}\, =\, \mathcal{L}_{\mathrm{QCD}}^{N_{f}=3}\, +\, [t_{L}]^{jl}_{ik}\; (\bar{q}_{L}{}^{i}\gamma^{\mu}q_{L}{}_j)\, (\bar{q}_{L}{}^{k}\gamma_{\mu}\, q_{L}{}_{l})\, +\, [t_{R}]^{jl}_{ik}\; (\bar{q}_{R}{}^{i}\gamma^{\mu} q_{R}{}_j)\, (\bar{q}_{R}{}^{k}\gamma_{\mu}\, q_{R}{}_{l}) \, ,
\end{equation}
with auxiliary tensorial sources $[t_{L,R}]^{jl}_{ik}$. These sources will be later identified with the corresponding Wilson coefficients of the short-distance electroweak Lagrangian, which are obviously scale and scheme dependent because the four-quark operators need to be properly renormalized.

Taking into account the transformation of the quark currents under $P$ and $C$,
\begin{equation}
(\bar{q}_{L(R)}{}^{i}\gamma^{\mu}q_{L(R)}{}_j)\,\overset{P}{\rightarrow}\, (\bar{q}_{R(L)}{}^{i}\gamma_{\mu}\, q_{R(L)}{}_j)\, , 
\qquad\qquad  
(\bar{q}_{L(R)}{}^{i}\gamma^{\mu} q_{L(R)}{}_j) \,\overset{C}{\rightarrow}\, -(\bar{q}_{R(L)}{}^{j}\gamma^{\mu} q_{R(L)}{}_i) \, ,\;
\end{equation}
invariance under $P$ and $C$ is recovered if
\begin{equation}\label{eq:discrete}
[t_{L(R)}]^{jl}_{ik}\,\overset{P}{\rightarrow}\, [t_{R(L)}]^{jl}_{ik}\, , 
\qquad\qquad\qquad
[t_{L(R)}]^{jl}_{ik}\,\overset{C}{\rightarrow}\, [t_{R(L)}]^{ik}_{jl} \, .
\end{equation}
Moreover, under chiral flavour transformations
\begin{equation}\label{eq:flavor}
[t_{L(R)}]^{jl}_{ik}\,\rightarrow\, L(R)^{\dagger\ j}_{j'} L(R)^{\dagger\ l}_{l'}\; t^{j'l'}_{i'k'}\; L(R)^{\ i'}_{i} L(R)^{\ k'}_{k}\ ,
\end{equation}
in order to preserve the chiral invariance of $\mathcal{L}$. Imposing this formal symmetry on the external sources (spurions), one can easily work out the symmetry implications for the different types of four-fermion operators.

It is convenient to identify those combinations of four-quark operators belonging to irreducible representations of the chiral group. The transformation (\ref{eq:flavor}) corresponds to the 81-dimensional representation $(\bar 3\otimes 3)\otimes (\bar 3\otimes 3)$ of $SU(3)_{L(R)}$, which can be decomposed into irreducible symmetric/antisymmetric representations with dimensions\footnote{The representation $r$ stands here for $(1_L,r_R)$ or $(r_L,1_R)$, corresponding to $[t_R]$ and $[t_L]$, respectively.} 1, 8, 10 and 27. This decomposition can be done in a straightforward way, taking into account that the $SU(3)$ transformations preserve traces and the symmetry under exchange of upper ($j\leftrightarrow l$) and/or lower ($i\leftrightarrow k$) indices.\footnote{
A pair of upper or lower indices give 6 symmetric plus 3 antisymmetric possibilities
($3\otimes 3 = 6 \oplus 3$). Considering the single and double traces of an upper and a lower index, the 36 symmetric-symmetric (SS) configurations can be decomposed in 27 ($=36-9$) fully traceless ones, plus other 8 ($=36-27-1$) configurations with non-vanishing single traces but null double trace, 
plus the singlet combination where both traces are non-zero~\cite{Lehner:2011fz}. Obviously, the 9 antisymmetric-antisymmetric (AA) configurations can only produce the octet plus singlet possibilities. A singlet combination cannot be present in the AS or SA configurations, which are then decomposed into $10\oplus 8$.
}

Defining a tensor scalar product as
\begin{equation}
\mathcal{A}\cdot \mathcal{B}\, =\, A^{ij}_{kl}\, B_{ij}^{kl}\, ,
\end{equation}
one can define an orthonormal basis 
in terms of irreducible subsets:
\begin{equation}
\{e^{a}_{27}{}_{S}^{S},e^{a}_{8}{}_{S}^{S},e^{}_{1}{}_{S}^{S},e^{a}_{10}{}_{S}^{A},e^{a}_{8}{}_{S}^{A},e^{a}_{10}{}_{A}^{S},e^{a}_{8}{}_{A}^{S},e^{a}_{8}{}_{A}^{A},e^{}_{1}{}_{A}^{A}\}\, ,
\end{equation}
where $S$ and $A$ refer to the symmetric or antisymmetric character of the representation with respect to the upper or lower indices. 
One can then write any tensor in this basis as
\begin{equation}
\mathcal{A}\, =\,  \mathcal{A}_{r}^{a}{}^{M}_{N} \ e_{r}^{a} {}^{M}_{N} \, ,
\end{equation}
where the coefficient is, using orthonormality,
\begin{equation}
\mathcal{A}_{r}^{a}{}^{M}_{N}
\, =\, \mathcal{A}\cdot e_{r}^{a} {}^{M}_{N}
\, =\, A^{ij}_{kl} \ \left[ e_{r}^{a} {}^{M}_{N}\right]^{kl}_{ij}\, .
\end{equation}

Since the operators in Eq.~(\ref{eq:sourllll}) are symmetric under the simultaneous exchange  $(i,k)\leftrightarrow (j,l)$, we only need to consider the symmetric-symmetric ($1\oplus 8\oplus 27$) and antisymmetric-antisymmetric ($1\oplus 8$) configurations.
The fully-symmetric singlet and octet basis elements take the form:
\begin{eqnarray}
\left[ e_{1}{}^{S}_{S}\right]_{ik}^{jl} & =&  \frac{1}{\sqrt{24}}\, (\delta_{i}^{j}\delta_{k}^{l}+\delta_{i}^{l}\delta_{k}^{j}) \, ,
\\
\left[ e^{a}_{8}{}_{S}^{S}\right]_{ik}^{jl} & =& \frac{1}{\sqrt{40}}\,
(\lambda_{i}^{a,j}\delta_{k}^{l}+\lambda_{i}^{a,l}\delta_{k}^{j}+\lambda_{k}^{a,j}\delta_{i}^{l} +\lambda_{k}^{a,l}\delta_{i}^{j}) \, ,
\label{eq:OctetSymBasis}
\end{eqnarray}
with $\lambda_{i}^{a,j}$ any basis of traceless $SU(3)$ matrices such that $\mathrm{Tr}(\lambda^{a}\lambda^{b}) = 2\delta^{ab}$, for which we adopt the conventional Gell-Mann choice.
Instead of building an explicit basis of 27 symmetric tensors, it is simpler to subtract the singlet and octet pieces from the symmetric-symmetric component of the tensor:
\begin{equation}\label{eq:27plet}
\mathcal{A}_{27}^{}{}^S_S\, =\, \mathcal{A}_{S}^{S} - (\mathcal{A}_{S}^{S}\cdot e_{1}{}^{S}_{S})\; e_{1}{}^{S}_{S} - (\mathcal{A}_{S}^{S}\cdot e^{a}_{8}{}^{S}_{S})\; e^{a}_{8}{}^{S}_{S} \, .
\end{equation}
The remaining antisymmetric-antisymmetric pieces can be projected in a fully analogous way with the corresponding basis elements
\begin{eqnarray}
\left[ e_{1}{}^{A}_{A}\right]_{ik}^{jl} & =&  \frac{1}{\sqrt{12}}\, (\delta_{i}^{j}\delta_{k}^{l}-\delta_{i}^{l}\delta_{k}^{j}) \, ,
\\
\left[ e^{a}_{8}{}_{A}^{A}\right]_{ik}^{jl} & =& \frac{1}{\sqrt{8}}\,
(\lambda_{i}^{a,j}\delta_{k}^{l}-\lambda_{i}^{a,l}\delta_{k}^{j}-\lambda_{k}^{a,j}\delta_{i}^{l}+\lambda_{k}^{a,l}\delta_{i}^{j}) \, .
\label{eq:OctetAsymBasis}
\end{eqnarray}

\subsection{\boldmath Effective \texorpdfstring{$\chi$}{ch}PT operators}

To build the corresponding structures in the low-energy $\chi$PT framework, one just needs to combine the transformation properties in Eqs.~(\ref{eq:discrete}) and (\ref{eq:flavor}) with those of the basic chiral building blocks. Under $P$ and $C$ \cite{Ecker:1988te},
\begin{eqnarray}
(D_{\mu_{1}}\cdots D_{\mu_{n}} U)_{i}^{j}\,\overset{P}{\rightarrow}\, (D^{\mu_{1}}\cdots D^{\mu_{n}} U)^{\dagger}{}
_{i}^{j} \, , 
\qquad && \qquad 
\chi_{i}^{j}\,\overset{P}{\rightarrow}\, \chi^{\dagger}{}_{i}^{j} \, ,
\\
(D_{\mu_{1}}\cdots D_{\mu_{n}} U)_{i}^{j}\,\overset{C}{\rightarrow}\, (D_{\mu_{1}}\cdots D_{\mu_{n}} U){}_{j}^{i} \, ,
\qquad && \qquad 
\chi_{i}^{j}\,\overset{C}{\rightarrow}\, \chi{}_{j}^{i} \, ,
\end{eqnarray}
and under flavour,\footnote{Schematically, for building blocks purposes one may just represent them as
$t^{LL}_{LL}, U^{L}_{R}, \chi^{L}_{R}, U^{\dagger}{}^{R}_{L}, \chi^{\dagger}{}^{R}_{L},t^{RR}_{RR}$.}
\begin{equation}
(D_{\mu_{1}}\cdots D_{\mu_{n}} U)_{i}^{j}\,\rightarrow\, R_{i}^{i'}\, (D_{\mu_{1}}\cdots D_{\mu_{n}} U)_{i'}^{j'}\, L_{j'}^{\dagger j} \, , 
\qquad\qquad 
\chi_{i}^{j}\,\rightarrow\, R_{i}^{i'}\,\chi_{i'}^{j'}\, L_{j'}^{\dagger j} \, .
\end{equation}
It turns useful to define simple $\chi$PT structures transforming as pure left or right objects:
\begin{eqnarray}
L_{\mu}\, \equiv\, i\, U^{\dagger}D_{\mu}U \,\rightarrow\, L\, L_\mu\, L^\dagger\, ,\hskip .5cm
\qquad &&\qquad
U^{\dagger}\chi\, \,\rightarrow\, L\, U^{\dagger}\chi\, L^\dagger\, ,
\\
R_{\mu}\, \equiv\, i\, U\, (D_{\mu}U)^{\dagger}\,\rightarrow\, R\, R_\mu\, R^\dagger\, ,
\qquad &&\qquad
U\,\chi^{\dagger} \,\rightarrow\, R\, U\,\chi^{\dagger} R^\dagger\, .
\end{eqnarray}
The LO building block compatible with a non-zero 27-plet arises at $\mathcal{O}(p^{2})$ from connecting $L_{\mu}{}_{k}^{i}L^{\mu}{}_{l}^{j}$ to $[t_{L}]^{kl}_{ij}$ and $R_{\mu}{}_{k}^{i}R^{\mu}{}_{l}^{j}$ to $[t_{R}]^{kl}_{ij}$. Making use of the relation~(\ref{eq:27plet}) to project the 27 piece, and requiring invariance under the discrete symmetries $P$ and $C$, one finds:
\begin{align}
\mathcal{L}_{27}\, =\, a_{27} \,\frac{F^4}{8}\, &\Bigg\{ [t_{L}]^{jl}_{ik}\;  \Big(  [L^{\mu,i}_{j}L_{\mu,l}^{k}+L^{\mu,k}_{j}L_{\mu,l}^{i}]
-\frac{1}{12}\,\mathrm{Tr}(L_{\mu}L^{\mu})\, [\delta^{i}_{j}\delta^{k}_{l}+\delta^{i}_{l}\delta^{k}_{j}] 
\nonumber\\
&\hskip 1.2cm\mbox{} -\frac{1}{10}\,\mathrm{Tr}(\lambda^{a}L_{\mu}L^{\mu})\, [\lambda_{a,j}^{i}\delta^{k}_{l}+\lambda_{a,l}^{i}\delta^{k}_{j}+\lambda_{a,j}^{k}\delta^{i}_{l}+\lambda_{a,l}^{k}\delta^{i}_{j}] \, +\, \mathcal{O}(p^{4}) \Big)
\nonumber\\ 
& +\,  L\leftrightarrow R\Bigg\}\, . 
\label{eq:27gen}
\end{align}
Parity invariance requires the result to be symmetric under the exchange $L\leftrightarrow R$. Therefore, the $(27_L,1_R)$ and $(1_L,27_R)$ components share the same normalization.
Symmetries alone do not allow to fix the ($\mu$ dependent) global constant $a_{27}$, which encodes details on the non-perturbative QCD dynamics. We have normalized it with a  factor $F^4$ so that $a_{27}$ is a dimensionless quantity.
Notice that there are no other independent colour or spinor structures that can give a $27$-plet made out of four-quark operators.\footnote{For ($V\pm A)\times(V \pm A)$ operators, Fierz transformations trivially relate the two possible colour-singlet structures.} At this chiral order, our non-perturbative ignorance for the 27-plet part of any (SM or beyond SM) effective four-quark operator is encoded in a single constant.

Projecting with the fully-symmetric octet basis element in Eq.~(\ref{eq:OctetSymBasis}), one directly finds the effective symmetric octet Lagrangian:\footnote{
An additional allowed octet structure is obtained, replacing $\mathrm{Tr}(\lambda^{a}L_{\mu}L^{\mu})$ and $\mathrm{Tr}(\lambda^{a}R_{\mu}R^{\mu})$ in Eq.~(\ref{eq:8sgen}) by $\mathrm{Tr}[\lambda^{a} (U^\dagger\chi +\chi^\dagger U)]$
and $\mathrm{Tr}[\lambda^{a} (U\chi^\dagger +\chi U^\dagger)]$, respectively \cite{Bernard:1985wf}.
However, it induces a vacuum misalignment through Goldstone tadpoles. Once the Goldstone fields are properly redefined, this $\mathcal{O}(p^2)$ weak mass term is rotated away. Thus, it does not contribute to any physical amplitude  \cite{Crewther:1985zt,Kambor:1989tz}. 
}
\begin{align}
\label{eq:8sgen}
\mathcal{L}_{8}^{S}\, &=\, a_{8}^{S}\, \frac{F^4}{80}\, \Big\{ [t_L]^{jl}_{ik}\,\mathrm{Tr}(\lambda^{a}L_{\mu}L^{\mu}) \, +\,
[t_R]^{jl}_{ik}\,\mathrm{Tr}(\lambda^{a}R_{\mu}R^{\mu}) \Big\}\, (\lambda_{a,j}^{i}\delta^{k}_{l}+\lambda_{a,l}^{i}\delta^{k}_{j}+\lambda_{a,j}^{k}\delta^{i}_{l}+\lambda_{a,l}^{k}\delta^{i}_{j})
\nonumber\\ &
+\,\mathcal{O}(p^{4}) \, .
\end{align}
A completely analogous derivation leads to the antisymmetric octet one:
\begin{align}
\label{eq:8agen}
\mathcal{L}_{8}^{A}\, &=\, - a_{8}^{A} \,\frac{F^4}{16}\, \Big\{ [t_L]^{jl}_{ik}\,\mathrm{Tr}(\lambda^{a}L_{\mu}L^{\mu})\, +\, [t_R]^{jl}_{ik}\,\mathrm{Tr}(\lambda^{a}R_{\mu}R^{\mu}) \Big\}\, (\lambda_{a,j}^{i}\delta^{k}_{l}-\lambda_{a,l}^{i}\delta^{k}_{j}-\lambda_{a,j}^{k}\delta^{i}_{l}+\lambda_{a,l}^{k}\delta^{i}_{j})
\nonumber\\ &
+\,\mathcal{O}(p^{4}) \, .
\end{align}

The parameters $a_{27}$, $a_8^S$ and  $a_8^A$ depend on the short-distance renormalization scale $\mu$. Since there is only a 27-plet structure, the $\mu$ dependence encoded in $a_{27}(\mu)$ cancels exactly the one carried by the $[t_L]$ and $[t_R]$ tensorial sources in Eq.~(\ref{eq:27gen}). The cancellation of renormalization-scale dependences is more subtle in the octet sector because the QCD interaction mixes different flavour-octet structures.

With only symmetry consideration, no  useful information can be derived from the singlet structures, since there are
pure $\mathcal{O}(p^{0})$ contact terms, such as $[t_{L(R)}]^{ij}_{ij}$ and $[t_{L(R)}]^{ij}_{ji}$, that are not related to the Goldstone dynamics.

\subsection{Left-right four-quark operators}

Let us now consider the Lagrangian
\begin{equation}\label{eq:sourlr}
\mathcal{L}\, =\,\mathcal{L}_{\mathrm{QCD}}^{N_{f}=3}\, +\, [t_{LR}^{\delta\delta}]^{jl}_{ik}\; (\bar{q}_{L}{}^{i}\gamma^{\mu}q_{L}{}_j)\, (\bar{q}_{R}{}^{k}\gamma_{\mu}q_{R}{}_{l})\, +\, [t_{LR}^{\lambda\lambda}]^{jl}_{ik}\; (\bar{q}_{L}{}^{i}\gamma^{\mu} T^a q_{L}{}_j)(\bar{q}_{R}{}^{k}\gamma_{\mu} T^{a} q_{R}{}_{l})\, ,
\end{equation}
where $T^a = \frac{1}{2}\, \lambda_C^a$ are the generators of the colour $SU(3)_C$ group with $\lambda_C^a$ the corresponding Gell-Mann matrices in colour space.
Both $t_{LR}^{\delta\delta}$ and $t_{LR}^{\lambda\lambda}$ share the same symmetry transformations. We will omit the superscript when we refer to any of them. The Lagrangian is invariant under discrete symmetries provided that
\begin{equation}\label{eq:LRdisc}
[t_{LR}]_{ik}^{jl}\,\overset{P}{\rightarrow}\, [t_{LR}]_{ki}^{lj} \, ,
\qquad\qquad  
[t_{LR}]_{ik}^{jl}\,\overset{C}{\rightarrow}\, [t_{LR}]_{lj}^{ki} \, ,
\end{equation}
while invariance under chiral flavour transformations requires
\begin{equation}\label{eq:flavorLR}
[t_{LR}]^{jl}_{ik}\,\rightarrow\, L^{\dagger\ j}_{j'}R^{\dagger\ l}_{l'}\; [t_{LR}]^{j'l'}_{i'k'}\; L^{\ i'}_{i} R^{\ k'}_{k}\, .
\end{equation}

The decomposition into irreducible representations is now simpler because each fermion bilinear transforms with a different $SU(3)$ group. Thus, we have the $\bar 3\otimes 3 = 1\oplus 8$ decomposition in each chiral sector, which results in four possible structures transforming as
$(1_{L},1_{R})$, $(8_{L},1_{R})$, $(1_{L},8_{R})$, and $(8_{L},8_{R})$. Following the same procedure explained before, an associated orthonormal basis is trivially given by
\begin{equation}
[e_{1_L,1_R}^{\phantom{a}}]_{ik}^{jl}  = \frac{1}{3}\delta_{i}^{j}\delta_{k}^{l}\, , \ \
[e^{a}_{8_L,1_R}]_{ik}^{jl}=\frac{1}{\sqrt{6}}\lambda^{a,j}_{L,i}\delta_{k}^{l}\, ,\ \
[e^{a}_{1_L,8_R}]_{ik}^{jl}=\frac{1}{\sqrt{6}}\delta_{i}^{j}\lambda^{a,l}_{R,k}\, , \ \ \label{eq:8x8bas}
[e^{ab}_{8_L,8_R}]_{ik}^{jl}=\frac{1}{2}\lambda^{a,j}_{L,i}\lambda^{b,l}_{R,k}\, ,
\end{equation}
where we have made explicit the left or right nature of the different Gell-Mann matrices.

The LO $\chi$PT structure compatible with a nonzero $(8_{L}, 8_{R})$ piece is the $\mathcal{O}(p^{0})$ tensor $U^{\dagger l}_{i} U_{k}^{j}$. Projecting it with the corresponding element of the orthonormal basis in Eq.~(\ref{eq:8x8bas}), one finds
\begin{equation}\label{eq:ew}
\mathcal{L}_{8_L, 8_R}^{\phantom{L}}\, =\, \frac{F^6}{4}\left( 
a^{\delta\delta}_{88}\, [t^{\delta\delta}_{LR}]_{ik}^{jl} \, +\,
a^{\lambda\lambda}_{88}\, [t^{\lambda\lambda}_{LR}]_{ik}^{jl} \right) \lambda_{L,j}^{a,i}\lambda_{R,l}^{b,k}\;
\mathrm{Tr}(\lambda^{a}_L U^{\dagger}\lambda^{b}_R U)
\, +\,\mathcal{O}(p^{2}) \, .
\end{equation}
It can be easily checked that this Lagrangian is invariant under parity and charge conjugation, provided that the external sources transform as indicated in Eq.~(\ref{eq:LRdisc}). 
These two discrete transformations connect the $(8_{L}, 1_{R})$ and $(1_{L}, 8_{R})$ sectors; their corresponding LO effective Lagrangian is easily found to be
\begin{align}\nonumber
\mathcal{L}_{8}^{LR}&=\frac{F^4}{6} \left( a^{\delta\delta}_{LR}\, [t^{\delta\delta}_{LR}]_{ik}^{jl}\, +\, a^{\lambda\lambda}_{LR}\, [t^{\lambda\lambda}_{LR}]_{ik}^{jl} \right)
\left\{\mathrm{Tr}(\lambda^{a}_L L_{\mu}L^{\mu})\; \lambda^{a,i}_{L,j}\delta_{l}^{k}\, +\, \mathrm{Tr}(\lambda^{a}_R R_{\mu}R^{\mu})\;\delta^{i}_{j}\lambda^{a,k}_{R,l}\right\} \, \\&+\, \mathcal{O}(p^4) \, .\label{eq:8lrgen}
\end{align}
The global factors $F^6$ and $F^4$ have been introduced in order to have dimensionless couplings $a^{\delta\delta}_{88}$, $a^{\lambda\lambda}_{88}$, $a^{\delta\delta}_{LR}$ and $a^{\lambda\lambda}_{LR}$.
Once again, the low-energy realization of the remaining singlet structure does not give any useful information.

\subsection{\boldmath Large-\texorpdfstring{$N_C$}{Nc} limit}
\label{subsec:LargeNc}

At LO in the momentum expansion, all non-trivial dynamical information about the non-singlet flavour structures is then encoded in the seven couplings $a_i(\mu)$. Their expected size can be easily estimated in the limit of a large number of QCD colours $N_C$, where the colour-singlet currents factorize. The LO $\chi$PT realizations of the left and right QCD currents are just given by \cite{Pich:1995bw}
\begin{equation}
(\bar{q}_{L}{}^{i}\gamma^{\mu}q_{L}{}_j)\,\dot=\, -\frac{1}{2}\,F^2\, L^{\mu,i}_j\, ,
\qquad\qquad
(\bar{q}_{R}{}^{i}\gamma^{\mu}q_{R}{}_j)\,\dot=\, -\frac{1}{2}\, F^2\, R^{\mu,i}_j\, .
\end{equation}
This explains the chosen normalization factor in Eq.~(\ref{eq:27gen}), from which the global factors in (\ref{eq:8sgen}) and (\ref{eq:8agen}) follow. Therefore, the dynamical couplings associated with left-left and right-right four-quark operators take the large-$N_C$ values:
\begin{equation}
\label{eq:a_iLL-LargeNC}
a_{27}^\infty(\mu) \, = \, 1\, ,
\qquad\qquad a_{8}^{S,\infty}(\mu) \, =\, 1\, ,
\qquad\qquad a_{8}^{A,\infty}(\mu) \, =\, 1\, .
\end{equation}

The left-right colour-singlet structure in Eq.~(\ref{eq:sourlr}) does not contribute to the LO $\chi$PT Lagrangians (\ref{eq:ew}) and (\ref{eq:8lrgen}) when $N_C\to\infty$:
\begin{equation}
\label{eq:a_iLRd-LargeNC}
a^{\delta\delta,\infty}_{88}(\mu) \, = \, 0\, ,   
\qquad\qquad 
a_{LR}^{\delta\delta,\infty}(\mu) \, = \, 0\, .
\end{equation}
Making a Fierz rearrangement, the colour-octet term can be written as a product of right and left scalar currents: 
\begin{align}\label{eq:LargeNcLR}
 (\bar{q}_{L}{}^{i}&\gamma^{\mu} T^a q_{L}{}_j)\, (\bar{q}_{R}{}^{k}\gamma_{\mu} T^a q_{R}{}_{l}) 
 \, =\,
 - (\bar{q}_{L}{}^{i} q_{R}{}_{l}) \, (\bar{q}_{R}{}^{k}q_{L}{}_j)
 +\frac{1}{N_C}\; (\bar{q}_{L}^{\,\alpha i} q_{R l}^{\, \beta}) \, (\bar{q}_{R}^{\,\beta k} q_{L j}^{\alpha})
 \nonumber\\  &\,\dot=\,  
 -B_0^2 F^2\,\left\{ \frac{1}{4}\, F^2\; U_l^i\, U^{\dagger k}_j 
 \right.\nonumber\\ &\left.
 +\,  U_l^i\,\left[ L_5\; U^\dagger D_\nu U D^\nu U^\dagger + 2 L_7\, U^\dagger\,\mathrm{Tr} (U^\dagger\chi-\chi^\dagger U) + 2 L_8\, U^\dagger\chi U^\dagger + H_2\,\chi^\dagger \right]_j^k
 \right.\nonumber\\ &\left.
 +\, \left[ L_5\; U D_\nu U^\dagger D^\nu U - 2 L_7\, U\,\mathrm{Tr} (U^\dagger\chi-\chi^\dagger U) + 2 L_8\, U\chi^\dagger U + H_2\,\chi \right]_l^i\,  U^{\dagger k}_j
\right\}\, ,
\nonumber\\ & +\, \mathcal{O}(p^4 N_C^2, p^2 N_C)\, ,
\end{align}
where the indices $\alpha$ and $\beta$ in the first line denote the quark colours (colour-singlet currents are understood whenever colour labels are not explicit).
In the last expression we have only kept the large-$N_C$ contributions, using the known $\chi$PT realization of these currents \cite{Gisbert:2017vvj}.
This fixes the normalization of $a^{\lambda\lambda}_{88}$ and $a^{\lambda\lambda}_{LR}$
in the limit $N_C\to\infty$:
\begin{equation}
\label{eq:a_iLRl-LargeNC}
a^{\lambda\lambda,\infty}_{88}(\mu) \, = \, - \frac{1}{4}\, B(\mu)\, ,
\qquad\qquad 
a_{LR}^{\lambda\lambda,\infty}(\mu) \, = \, -2\, L_5\, B(\mu)\, ,
\end{equation}
where the $\mu$-dependent factor
\begin{align}\label{eq:Bmu}
B(\mu)\,\equiv\, \frac{B^2_0}{F^2}
\, =\, \left[ \frac{M_K^2}{(m_s+m_d)(\mu)\, F_\pi}\right]^2 
&\left\{1+\frac{8 M_\pi^2}{F_\pi^2}\, L_5
-\frac{16 M_K^2}{F_\pi^2}\, (2 L_8 - L_5)
\right\}
\end{align}
is related to the quark condensate in the chiral limit,
$\langle 0|\bar u u|0\rangle = -F^2 B_0$. The constants $L_i$ and $H_2$ are low-energy couplings of the $\mathcal{O}(p^4)$ $\chi$PT Lagrangian \cite{Pich:1995bw}. 

Thus,  $F^2 a^{\lambda\lambda}_{88}(\mu)$ and $a_{LR}^{\lambda\lambda}(\mu)$ are of $\mathrm{O}(N_C^0)$, while $a^{\delta\delta}_{88}(\mu)$ and $a_{LR}^{\delta\delta}(\mu)$ are suppressed by a factor $1/N_C$. The dependence on the renormalization scale of $a_{27}(\mu)$, $a_{8}^S(\mu)$ and $a_{8}^A(\mu)$ is also colour suppressed, while the factor $B(\mu)$ captures the exact $\mu$ dependence of $a^{\lambda\lambda,\infty}_{88}(\mu)$ and  $a_{LR}^{\lambda\lambda,\infty}(\mu)$ in the large-$N_C$ limit. The anomalous dimensions of the left-left and right-right operators are necessarily of NLO in $1/N_C$ because the vector and axial-vector currents are not renormalized. On the other side, the scalar and pseudoscalar QCD currents do depend on renormalization conventions. Only renormalization-invariant combinations such as $m_q \,\bar q^i q_j$ can appear in observable quantities, which explains  why the $\mu$ dependence of left-right operators scales with the factor $B(\mu)\sim m_q(\mu)^{-2}$ at large-$N_C$.

\section{Strangeness-changing weak transitions}
\label{sec:DStransitions}

Let us particularize now the previous discussion to the $\Delta S=1$ and $\Delta S=2$ transitions.
After integrating out the heavy mass scales, the effective $\Delta S=1$ SM Lagrangian takes the form
\begin{equation}\label{eq:efflag}
\mathcal{L}^{\Delta S=1}\, =\, G\,\sum_{i=1}^{10} C_{i}(\mu)\; Q_{i}(\mu)\, ,
\end{equation}
where 
\begin{equation}\label{eq:Gdef}
G\,\equiv\, -\frac{G_{F}}{\sqrt{2}}\, V_{ud}^{\phantom{*}} V_{us}^{*}\, ,
\end{equation}
contains the Fermi coupling and the leading quark-mixing parameters, and the sum extends over the standard basis of ten four-quark operators $Q_{i}$~\cite{Buchalla:1995vs,Gilman:1979bc}: 
\begin{align}
    Q_1 &= 4\, (\bar s_L^{\alpha}\gamma^\mu u_L^{\beta})\, (\bar u_L^{\beta}\gamma_\mu d_L^{\alpha})\, ,
    &
    Q_2 &= 4\, (\bar s_L\gamma^\mu u_L)\, (\bar u_L\gamma_\mu d_L)\, ,
    \nonumber\\
    Q_3 &= 4\, (\bar s_L\gamma^\mu d_L)\,\sum_{q=u,d,s} (\bar q_L\gamma_\mu q_L)\, ,
    & 
    Q_4 &= 4\,  (\bar s_L^\alpha\gamma^\mu d_L^\beta)\,\sum_{q=u,d,s} (\bar q_L^\beta\gamma_\mu q_L^\alpha)\, ,
    \nonumber\\
    Q_5 &= 4\, (\bar s_L\gamma^\mu d_L)\,\sum_{q=u,d,s} (\bar q_R\gamma_\mu q_R)\, ,
    & 
    Q_6 &= 4\, (\bar s_L^\alpha\gamma^\mu d_L^\beta)\,\sum_{q=u,d,s} (\bar q_R^\beta\gamma_\mu q_R^\alpha)\, ,
    \nonumber\\
    Q_7 &= 6\, (\bar s_L\gamma^\mu d_L)\,\sum_{q=u,d,s} e_q\, (\bar q_R\gamma_\mu q_R)\, ,
    & 
    Q_8 &= 6\, (\bar s_L^\alpha\gamma^\mu d_L^\beta)\,\sum_{q=u,d,s} e_q\, (\bar q_R^\beta\gamma_\mu q_R^\alpha)\, ,
 \nonumber\\
    Q_9 &= 6\, (\bar s_L\gamma^\mu d_L)\,\sum_{q=u,d,s} e_q\, (\bar q_L\gamma_\mu q_L)\, ,
    & 
    Q_{10} &= 6\, (\bar s_L^\alpha\gamma^\mu d_L^\beta)\,\sum_{q=u,d,s} e_q\, (\bar q_L^\beta\gamma_\mu q_L^\alpha)\, ,
\label{eq:DS-4quark-Operators}
\end{align}
where $\alpha,\beta$ are colour indices.
The factors $e_q$ denote the corresponding quark charges in units of $e=\sqrt{4\pi\alpha}$.
All short-distance dynamical information on the heavy scales is encoded in the Wilson Coefficients $C_{i}(\mu)=z_{i}(\mu)+\tau\, y_{i}(\mu)$, where $\tau = -\lambda_t/\lambda_u$ with $\lambda_q = V_{qs}^* V_{qd}^{\phantom{*}}$. These coefficients can be computed with standard perturbative tools and their numerical values at NLO
are given in Table \ref{tab:wil}. 

\begin{table}[t]\centering
\begin{tabular}{|l|l|l|}
\hline
   & $z_{i}(1\,\mathrm{GeV})$ & $y_{i}(1\,\mathrm{GeV})$\\ \hline
$1$  & $-0.482$                           & $0$                                \\ \hline
$2$  & $1.260$                             & $0$                                \\ \hline
$3$  & $0.00105$                          & $0.0307$                           \\ \hline
$4$  & $-0.0296$                          & $-0.0563$                          \\ \hline
$5$  & $0.00699$                          & $0.00105$                          \\ \hline
$6$  & $-0.0293$                          & $-0.103$                          \\ \hline
$7$  & $0.0000745$                        & $-0.000314$                        \\ \hline
$8$  & $0.0000832$                        & $0.00115$                          \\ \hline
$9$  & $0.000117$                        & $-0.0114$                          \\ \hline
$10$ & $-0.0000503$                       & $0.00475$                          \\ \hline
\end{tabular}
\caption{\label{tab:wil}$\Delta S=1$ Wilson Coefficients at NLO in the $\overline{\mathrm{MS}}$ (NDR) scheme at $\mu=1$~GeV.}
\end{table}

The effective realization of $\mathcal{L}^{\Delta S=1}$ in the low-energy Goldstone theory is well known~\cite{Cirigliano:2011ny}. At LO is characterized by three different $\chi$PT structures \cite{Cronin:1967jq,Bijnens:1983ye,Bernard:1985wf,Grinstein:1985ut,Guberina:1985md,Pich:1985st,Kambor:1989tz,Pich:1990mw},
\begin{equation}
\label{eq:LO_DS_ChPT_L}
 \mathcal{L}^{\Delta S=1}_{\mathrm{eff}} =  G F^4\left\{ 
 g_{27}\, \Bigl( L^{\mu 3}_2 L_{\mu 1}^1 + \frac{2}{3}\, L^{\mu 1}_2 L_{\mu 1}^3\Bigr)
 + g_8\, \mathrm{Tr}(\lambda L^\mu L_\mu)    
 + e^2 g_8\, g_{\mathrm{ewk}} \, F^2\, \mathrm{Tr}(\lambda U^\dagger Q U)
 \right\}  ,
\end{equation}
transforming as $(27_L,1_R)$, $(8_L,1_R)$ and $(8_L,8_R)$, respectively. Here, 
$\lambda\equiv\frac{1}{2}\, (\lambda_{6}-i\lambda_{7})$ projects onto the $\bar s\to \bar d$ transition and $Q= \frac{1}{3}\, \mathrm{diag}(2,-1-1)$ is the quark charge matrix.

Particularizing the tensor sources in Eqs.~(\ref{eq:sourllll}) and (\ref{eq:sourlr}) to the SM $\Delta S=1$ Lagrangian (\ref{eq:efflag}) and projecting over the different chiral-symmetry components, using Eqs.~(\ref{eq:27gen}), (\ref{eq:8sgen}), 
(\ref{eq:8agen}), (\ref{eq:ew}) and (\ref{eq:8lrgen}), one easily finds the expression of the three low-energy couplings in terms of the SM Wilson coefficients:\footnote{
The operator basis is redundant because $Q_4 = -Q_1 + Q_2 +Q_3$, \ $Q_9 = \frac{3}{2}\, Q_1 - \frac{1}{2}\, Q_3$ \ and \
$Q_{10} = \frac{1}{2}\, Q_1 +Q_2 - \frac{1}{2}\, Q_3$. Thus, $Q_4$, $Q_9$ and $Q_{10}$ can be eliminated redefining appropriately the Wilson coefficients to $C_1' = C_1-C_4 + \frac{3}{2}\, C_9 + \frac{1}{2}\, C_{10}$, \ $C_2' = C_2+C_4 + C_{10}$ \ and \ $C_3' = C_3+C_4 - \frac{1}{2}\, C_9 - \frac{1}{2}\, C_{10}$.  The Fierz transformation needed to rewrite $Q_{4}$ in the colour-singlet form of Eq. (4) generates an additional tiny contribution from evanescent operators in the NDR scheme (this correction is zero with the 't Hooft-Veltman prescription for $\gamma_5$) \cite{Buras:1993dy}. It can be easily incorporated in Eq.~(\ref{eq:g8}) with the changes: 
$C_{6}\rightarrow C_{6}-\frac{\alpha_{s}}{4\pi}\, C_{4}$, \
$C_{4}\rightarrow C_{4}-\frac{\alpha_{s}}{4\pi}\, C_{4}$, \
$C_{3}\rightarrow C_{3}+\frac{\alpha_{s}}{12\pi}\, C_{4}$ \ and \
$C_{5}\rightarrow C_{5}+\frac{\alpha_{s}}{12\pi}\, C_{4}$.
\label{foot:OpRel}
}
%
\begin{eqnarray}\label{eq:g27}
g_{27}&=& \frac{3}{5}\, a_{27}(\mu)\; \Big( C_{1} +C_{2} +\frac{3}{2}\, C_{9} +\frac{3}{2}\, C_{10}\Big) (\mu) \, ,
\\
g_{8}&=& 
\frac{1}{10}\, a_{8}^{S}(\mu)\left(C_{1}+C_{2}+5\, C_{3}+5\, C_{4}-C_{9}-C_{10}\right)\! (\mu) 
\nonumber\\ 
&&
-\frac{1}{2}\, a_{8}^{A}(\mu) \left(C_{1}-C_{2}+C_{3}-C_{4}+C_{9}-C_{10}\right)\! (\mu)
\nonumber\\
&&
+ 4 \, a_{LR}^{\delta\delta}(\mu) \left(C_{5}+\frac{C_{6}}{N_{c}}\right)\! (\mu) + 8\, a_{LR}^{\lambda\lambda}(\mu)\, C_{6}(\mu)
\, , 
\label{eq:g8}
\\ 
e^2g_{8}\,g_{\mathrm{ewk}}&=& 6\left\{ a^{\delta\delta}_{88}(\mu) \left(C_{7}+\frac{C_{8}}{N_{c}}\right)\! (\mu)  
+2\, a^{\lambda\lambda}_{88}(\mu)\, C_{8}(\mu)  \right\} \, . 
\label{eq:g8gew}
\end{eqnarray}
Since the chiral couplings $g_{27}$, $g_8$ and $g_{\mathrm{ewk}}$ are independent of the short-distance renormalization scale $\mu$, these equations contain also information on the $\mu$ dependence of the non-perturbative parameters $a_i(\mu)$.
Inserting the large-$N_C$ values of the $a_i(\mu)$ couplings in Eqs.~(\ref{eq:a_iLL-LargeNC}), (\ref{eq:a_iLRd-LargeNC}) and (\ref{eq:a_iLRl-LargeNC}), one recovers the known expressions for the weak $\chi$PT LECs in the limit of a large number of QCD colours \cite{Pallante:2001he}:

\begin{eqnarray}\label{eq:g27g8gewLargaNc}
g_{27}^\infty &=& \frac{3}{5}\; \Big( C_{1}  +C_{2}  +\frac{3}{2}\, C_{9}  +\frac{3}{2}\, C_{10} \Big) \, ,
\nonumber\\
g_{8}^\infty &=& 
-\frac{2}{5}\, C_{1} + \frac{3}{5}\, C_{2} + C_{4} - 16\, L_5\, B(\mu)\, C_{6}(\mu)
- \frac{3}{5}\, C_{9} + \frac{2}{5}\, C_{10} \, ,
\nonumber\\ 
\left(e^2g_{8}\,g_{\mathrm{ewk}}\right)^\infty &=& 
-3\, B(\mu)\, C_{8}(\mu)   \, . 
\end{eqnarray}
$C_6$ and $C_8$ are the only Wilson coefficients carrying an explicit dependence on $\mu$ at $N_C\to\infty$. This dependence is exactly cancelled by the factor $B(\mu)$.

\subsection{\boldmath \texorpdfstring{$\Delta S =2$}{DS=2} Lagrangian}

In the SM the mixing between the neutral kaon and its antiparticle is mediated by box diagrams with two $W$ exchanges.
In the three-flavour theory, they generate a $\Delta S=2$ effective Lagrangian that contains one single dimension-six operator \cite{Gilman:1982ap}:
\begin{equation}
\mathcal{L}^{\Delta S=2}\, =\, -\frac{G_F^2M_W^2}{(4\pi)^2}\, \mathcal{F}(V_{\mathrm{CKM}},m_c,m_t)\; C_{\Delta S=2}(\mu)\; Q_{\Delta S=2}(\mu) \, ,
\end{equation}
where
\begin{equation}
Q_{\Delta S=2}\, =\, 4\, (\bar{s}_{L}\gamma_{\mu}d_L) \, (\bar{s}_L\gamma^{\mu}d_L) \, ,
\end{equation}
\begin{equation}
C_{\Delta S=2}(\mu)\, =\, \alpha_s(\mu)^{-2/9} \left[1 + \frac{\alpha_s(\mu)}{4\pi}\, J_3\right]    
\label{eq:C_DS2}
\end{equation}
and the short-distance factor \cite{Brod:2019rzc}
\begin{equation}
\mathcal{F}(V_{\mathrm{CKM}},m_c,m_t)\, =\, \lambda_u^2\,\eta_{cc}\, \mathscr{S}(x_c) +\lambda_t^2\,\eta_{tt}\, \mathscr{S}(x_t) + 2\lambda_u\lambda_t\,\eta_{ut}\, \mathscr{S}(x_c,x_t) 
\end{equation}
contains the information on the relevant quark-mixing factors $\lambda_q$
and the heavy mass scales, through the modified Inalmi-Lim functions $\mathscr{S}(x_q)$ and $\mathscr{S}(x_c,x_t)$, where $x_q=m_q^2/M_W^2$. In the $\overline{\mathrm{MS}}$ scheme, the QCD corrections take the values $J_3=1.895$, $\eta_{cc}=1.87\pm 0.76$, $\eta_{tt}=0.5765\pm 0.0065$ and $\eta_{ut}=0.402\pm 0.005$ \cite{Buras:1990fn,Herrlich:1993yv,Herrlich:1996vf,Brod:2010mj,Brod:2019rzc}. 

The corresponding external source tensor in Eq.~(\ref{eq:sourllll}), $[t_L]_{33}^{22}$, belongs to the $(27_L,1_R)$ multiplet. Using Eq.~(\ref{eq:27gen}), one finds the effective $\chi$PT realization of this $\Delta S=2$ operator:
\begin{align}\label{eq:DS=2eff}
\mathcal{L}_{27}^{\Delta S=2}\, =\, \frac{G_F^2M_W^2}{(4\pi)^2}\, \mathcal{F}(V_{\mathrm{CKM}},m_c,m_t)\; g_{\Delta S=2}\, F^4\; L^{\mu,3}_{2} L_{\mu,2}^{3} +\mathcal{O}(p^4)\, ,
\end{align}
with
\begin{equation}
g_{\Delta S=2}  \, =\,  
a_{27}(\mu)\, C_{\Delta S=2}(\mu) \, .
\end{equation}
Thus, $a_{27}(\mu)$ depends on the renormalization scale in precisely the opposite way than Eq.~(\ref{eq:C_DS2}), so that the product $g_{\Delta S=2}$ remains scale invariant.

Since both involve the same non-perturbative parameter $a_{27}(\mu)$, the chiral couplings $g_{\Delta S=2}$ and $g_{27}$ are directly related through the identity
\begin{equation}\label{eq:27relation}
g_{27}\, =\, \frac{3}{5}\, \frac{\Big( C_{1} +C_{2} +\frac{3}{2}\, C_{9} +\frac{3}{2}\, C_{10}\Big) (\mu)}{C_{\Delta S=2}(\mu)}\;   g_{\Delta S=2}
\, .  
\end{equation}
This symmetry relation guarantees that the running of the Wilson coefficients in the numerator matches exactly the one of $C_{\Delta S=2}(\mu)$ in the denominator, so that the ratio is scale invariant.
From the measured $K\to\pi\pi$ rates, one obtains at NLO in $\chi$PT \cite{Cirigliano:2011ny,Cirigliano:2019cpi}
\begin{equation}\label{eq:g27-exp}
g_{27}\,  =\, 0.29\pm 0.02\, ,
\end{equation} 
which implies
\begin{equation}\label{eq:gDS2-exp} 
g_{\Delta S=2}\, =\, 0.79\pm 0.05   \, ,
\end{equation} 
and
\begin{equation}\label{eq:a27-exp}
a_{27}(\mu_0)\, =\, 0.622 \pm 0.043 
\end{equation}
at  $\mu_0 = 1$~GeV.

The relation between these two 27-plet couplings is usually expressed \cite{Donoghue:1982cq} in terms of the so-called $B_K$ parameter, defined through
\begin{equation}
\langle \bar{K}^{0}| Q_{\Delta S=2} |K^{0}\rangle\, =\, \frac{16}{3}\, F_K^2 M_K^2\, B_K \, ,
\end{equation}
or the scale-invariant quantity
$\hat B_K \equiv B_K(\mu)\,  C_{\Delta S=2}(\mu)$.
Evaluating this hadronic matrix element with the effective Lagrangian (\ref{eq:DS=2eff}), one gets 
\begin{equation}
F_K^2 \hat B_K\, =\, \frac{3}{4}\, F^2\, g_{\Delta S=2} + \mathcal{O}(p^4)\, .
\end{equation}
Thus,  $\frac{3}{4} \, g_{\Delta S=2}$ and $\frac{3}{4} \, a_{27}(\mu)$ correspond to the values of $\hat B_K$  and $B_K(\mu)$, respectively, in the chiral limit.\footnote{
$B_K(\mu)$ receives large chiral corrections of $\mathrm{O}[M_K^2 \log{(M_K^2/\nu^2_\chi)}/\Lambda_\chi^2]$ \cite{Bijnens:1984ec}.}
Using Eq.~(\ref{eq:a_iLL-LargeNC}), one recovers the well-known result $B_K^\infty = \frac{3}{4}$ at large $N_C$.

The value of $g_{\Delta S=2}$ extracted above from the $K\to\pi\pi$ rates implies $\hat B_K = 0.59 \pm 0.02$ in the chiral limit. This can be compared with the results from explicit calculations with different methods:
\begin{equation}
\lim_{m_q\to 0}\, \hat B_K\, =\, \left\{ \begin{array}{ccc}
 0.33\pm 0.09    &  \text{\cite{Pich:1985ab}}\\
 0.38\pm 0.15    &  \text{\cite{Bijnens:2006mr,Bijnens:1995br}}\\
 0.36\pm 0.15    &  \text{\cite{Peris:2000sw,Cata:2003mn}}
\end{array}   
\right. .
\end{equation}
Conversely, taking the chiral-limit value of $\hat B_K$ from the most recent calculation of Ref.~\cite{Bijnens:2006mr}, one predicts:
\begin{equation}
\label{eq:g27pred}
g_{\Delta S=2}\, =\, 0.51 \pm 0.20\, ,
\qquad\qquad\qquad
g_{27}\, =\, 0.19\pm 0.07\, ,
\end{equation}
and
\begin{equation}
\label{eq:a27det}
a_{27}(\mu_0)\, =\, 0.40\pm 0.16 \, ,
\end{equation}
at $\mu_0 = 1$~GeV.

Since $a_{27}(\mu)$ is a CP-conserving parameter, Eq.~(\ref{eq:g27}) allows us to predict also the tiny CP-violating component of $g_{27}$. 
Taking the experimental value of $\mathrm{Re} [g_{27}]$ in Eq.~(\ref{eq:g27-exp}), one gets
\begin{equation}
    \mathrm{Im}[ g_{27}]\, = \,\frac{\mathrm{Im}\Big( C_{1} +C_{2} +\frac{3}{2}\, C_{9} +\frac{3}{2}\, C_{10}\Big)}{\mathrm{Re}\Big( C_{1} +C_{2} +\frac{3}{2}\, C_{9} +\frac{3}{2}\, C_{10}\Big)}\;  \mathrm{Re} [g_{27}]
\, = \,  -(0.0037\pm 0.0002)\;\mathrm{Im} (\tau)\, ,
\end{equation}
where $\mathrm{Im}(\tau)\approx -\eta \lambda^4 A^2/\sqrt{1-\lambda^2}$ in the Wolfenstein parametrization of the CKM matrix.

\section{Vacuum condensates}
\label{sec:condensates}

The two-point correlation functions of the colour-singlet vector $V^\mu_{ij} = \bar q^j\gamma^\mu q_i$ 
and axial-vector $A^\mu_{ij} = \bar q^j\gamma^\mu\gamma_5 q_i$ quark currents,
\begin{equation}\label{eq:Correlators-def}
\Pi^{\mu\nu}_{ij,\mathcal{J}}(q)\equiv i \int d^{4}x\; e^{iqx}\, \langle 0 |  T(\mathcal{J}_{ij}^{\mu}(x) \mathcal{J}_{ij}^{\nu}(0)^\dagger) |0\rangle=
(-g^{\mu\nu}q^{2}+q^{\mu}q^{\nu})\,\Pi^{L+T}_{ij,\mathcal{J}}(q^{2})+g^{\mu\nu}q^2\,\Pi^{L}_{ij,\mathcal{J}}(q^{2}) \, ,
\end{equation}
play a central role in the study of hadronic production through electroweak currents \cite{Pich:2020gzz}.
Here, $\mathcal{J}=V,A$ and the superscripts denote the transverse ($T$) and longitudinal ($L$) components. 
We are mainly interested in the correlators associated with $\mathcal{J}^\mu_{ud}$, $(V+A)^\mu_{us}$ and 
$\bar J^\mu_{\mathrm{em}} \equiv   \sqrt{3/2}             %
\,\sum_i e_i V^\mu_{ii}$, which can be related to precise experimental data.
From now on, we focus on their corresponding $L+T$ parts (omitting the $L+T$ label), which we will denote
$\Pi^{d}_{\mathcal{J}}$, $\Pi^{s}_{V+A}$ and $\overline{\Pi}_{EM}$.

At large Euclidean momenta $Q^{2}=-q^{2}\gg \Lambda^2_{\mathrm{QCD}}$, their asymptotic behaviour is well described by the OPE \cite{Shifman:1978bx}:
\begin{equation}\label{eq:wilsons}
\Pi(q^{2})\, =\, \sum_{i,D} \frac{C_{i,D}(q^{2},\mu)\,\langle\mathcal{O}_{i,D}(\mu)\rangle}{(-q^2)^{D/2}} 
\,\equiv\, \sum_{D} \frac{\langle\mathcal{O}_{D}\rangle}{(-q^2)^{D/2}} \, .
\end{equation}
The leading $D=0$ perturbative contribution, which is currently known to order $\alpha_{s}^{4}$ \cite{Baikov:2008jh,Baikov:2010je,Herzog:2017dtz,Baikov:2012zn},
is corrected by inverse-power contributions from gauge- and Lorentz-invariant operators of increasing dimension $D$.
These dimensional corrections, obtained by dressing and renormalizing contributions where not all quark and gluon fields are contracted, are characterized by Wilson coefficients that only depend logarithmically on the energy scale,
\begin{equation}
C_{i,D}(q^{2},\mu)\, =\, C_{i,D}^{0} \left\{ 1 + \alpha_{s}(\mu) \left[ c_{i,D} + c_{i,D}^L\, \log{(-q^2/\mu^2)}\right] +\mathcal{O}(\alpha_{s}^2)\right\} \, ,
\end{equation}
where the coefficients $c_{i,D}^L$ are related to the leading anomalous-dimension matrix of the associated operators.

We are going to analyze the four-quark operators that appear at $D=6$. Following a notation close to Eqs.~(\ref{eq:sourllll}) and (\ref{eq:sourlr}), their contributions to the relevant current correlators~\cite{Braaten:1991qm} can be written in the form\footnote{
For the left-left and right-right operators, the notation of Eq.~(\ref{eq:sourllll}) without colour matrices corresponds to $[t_{L(R)}]^{jl}_{ik} = \frac{1}{2}\left([\tilde t_{L(R)}]^{lj}_{ik} - \frac{1}{N_C}\, [\tilde t_{L(R)}]^{jl}_{ik}\right)$.}
%
\begin{align}
\mathcal{O}_6 \, &=\, 
[\tilde t_{L}]^{jl}_{ik}\; (\bar{q}_{L}^{i} \gamma^{\mu} T^a q_{Lj})\, (\bar{q}_{L}^{k} \gamma_{\mu} T^a q_{Ll}) 
\, +\, 
[\tilde t_{R}]^{jl}_{ik} \; (\bar{q}_{R}^{i} \gamma^{\mu} T^a q_{Rj})\, (\bar{q}_{R}^{k} \gamma_{\mu} T^a q_{Rl}) 
\nonumber\\[3pt] &
+\, 
[t_{LR}^{\delta\delta}]^{jl}_{ik}\;
(\bar{q}_{L}^{i} \gamma^{\mu} q_{Lj})\, (\bar{q}_{R}^{k}\gamma_{\mu}  q_{Rl})
+ [t_{LR}^{\lambda\lambda}]^{jl}_{ik}\;
(\bar{q}_{L}^{i} \gamma^{\mu} T^a q_{Lj})\, (\bar{q}_{R}^{k}\gamma_{\mu} T^a q_{Rl})
\, .
\end{align}
At LO in $\alpha_s$, $[t_{LR}^{\delta\delta}]^{jl}_{ik}=0$.
For $\Pi^{d}_{\mathcal{J}}$ the non-zero tensor coefficients are
\begin{align}\label{eq:tildet-d}
[\tilde t^{d}_{L,\mathcal{J}}]^{jl}_{ik}\, &=\, [\tilde t^{d}_{R,\mathcal{J}}]^{jl}_{ik}\, =\, 
-8\pi \alpha_{s} \left\{
\frac{1}{4}\, (\lambda_{i}^{1,j}\lambda_{k}^{1,l}+\lambda_{i}^{2,j}\lambda_{k}^{2,l}) 
+\frac{1}{18\sqrt{3}}\, (\lambda_{i}^{8,j}\delta_{k}^{l}+\lambda_{k}^{8,l}\delta_{i}^{j})
+ \frac{2}{27}\,\delta_{i}^{j}\delta_{k}^{l}
\right\} , 
\\ \label{eq:tildet-d-LR}
[t^{d,\lambda\lambda}_{LR,\mathcal{J}}]^{jl}_{ik}\, &=\,
-8\pi \alpha_{s} \left\{
\mp\frac{1}{2}\, (\lambda_{i}^{1,j}\lambda_{k}^{1,l}+\lambda_{i}^{2,j}\lambda_{k}^{2,l}) 
+\frac{1}{9\sqrt{3}}\, (\lambda_{i}^{8,j}\delta_{k}^{l}+\lambda_{k}^{8,l}\delta_{i}^{j})
+ \frac{4}{27}\, \delta_{i}^{j}\delta_{k}^{l}  
\right\} ,
\end{align}
where the upper (lower) signs correspond to the vector (axial-vector) currents. To obtain the corresponding results for the $\Pi^{s}_{\mathcal{J}}$ correlators, one just needs to exchange the down and strange quarks, which amounts to the changes
\begin{equation}
 [t^{s}_{\mathcal{J}}]^{jl}_{ik}\, =\, [t^{d}_{\mathcal{J}}]^{jl}_{ik}
 \Big(
 \lambda_1\rightarrow \lambda_4, \lambda_2\rightarrow \lambda_5,\lambda_8\rightarrow \frac{\sqrt{3}}{2}(\lambda_{3}-\frac{1}{\sqrt{3}}\lambda_8)
 \Big)\, .
\end{equation}
Finally, the tensor coefficients of the electromagnetic correlator $\overline{\Pi}_{EM}$ are
\begin{align}
[\tilde t^{}_{L,EM}]^{jl}_{ik}\, &=\, [\tilde t^{}_{R,EM}]^{jl}_{ik}\, =\, 
-8\pi \alpha_{s} \left\{
\frac{3}{2}\, Q_{i}^{j}Q_{k}^{l}
+\frac{1}{18}\, (Q_{i}^{j}\delta_{k}^{l}+\delta_{i}^{j}Q_{k}^{l})
+\frac{2}{27}\,\delta_{i}^{j}\delta_{k}^{l} \right\} , 
\\
[t^{\lambda\lambda}_{LR,EM}]^{jl}_{ik}\, &=\, -8\pi \alpha_{s} \left\{
-3\, Q_{i}^{j}Q_{k}^{l} 
+\frac{1}{9}\, (Q_{i}^{j}\delta_{k}^{l}+\delta_{i}^{j}Q_{k}^{l})  
+ \frac{4}{27}\,\delta_{i}^{j}\delta_{k}^{l}
\right\} .
\end{align}
In addition to the octet and 27-plet structures, all these correlators contain also flavour-singlet components. However, the singlet terms cancel in the flavour-breaking differences $\Pi^{d}_{V-A}$, $\Pi_{V+A}^{d-s}$ and $\overline{\Pi}_{EM} - \Pi^{d}_{V}$, together with the purely perturbative contributions.\footnote{
The so-called singlet topologies that only contribute to the neutral correlators are absent in the three-flavour theory because $\sum_q e_q = 0$ \cite{Pich:2020gzz}.
}
These correlation functions are then governed by long-distance matrix elements that can be related to the ones discussed in the previous section.

\subsection{\boldmath \texorpdfstring{$\mathcal{O}^{d}_{6,V-A}$}{O6V-A}}

The cleanest flavour-breaking difference is $\mathcal{O}^{d}_{6,V-A}$, which only receives an $(8_{L}, 8_{R})$ contribution from $[t_{LR,V-A}^{d,\lambda\lambda}]^{jl}_{ik}= 8\pi \alpha_{s}\, (\lambda_{L,i}^{1,j}\lambda_{R,k}^{1,l}+\lambda_{L,i}^{2,j}\lambda_{R,k}^{2,l}) $. 
From Eqs.~(\ref{eq:sourlr}) and (\ref{eq:ew}), the realization of this local operator in terms of the long-distance degrees of freedom is found to be:
\begin{equation}\label{eq:OV-A_Goldstone}
\mathcal{O}^{d}_{6,V-A}(\mu)\, =\,  8\pi \alpha_{s}(\mu)\, F^{6} a_{88}^{\lambda\lambda}(\mu)\; \mathrm{Tr}(\lambda^{1}_L U^{\dagger}\lambda_R^{1} U + \lambda^{2}_L U^{\dagger}\lambda^{2}_R U) + 
\mathcal{O}(p^2,\alpha_{s}^{2})\, .
\end{equation}
Taking now the vacuum expectation value, one finds
\begin{equation}\label{eq:OV-A_cond}
\langle\mathcal{O}^{d}_{6,V-A}(\mu)\rangle\, =\, 32\pi \alpha_{s}(\mu)\, F^{6}a_{88}^{\lambda\lambda}(\mu) 
+ \mathcal{O}(p^2,\alpha_{s}^{2})\, ,
\end{equation}
which provides a direct link between this condensate and $g_{8}g_{\mathrm{ewk}}$ in Eq.~(\ref{eq:g8gew}). 

Expanding the flavour trace in Eq.~(\ref{eq:OV-A_Goldstone}) to second order in the Goldstone fields and computing the resulting tadpole contributions, we can easily obtain the $\mathcal{O}(p^2)$ $\chi$PT corrections to the vacuum condensate:
\begin{align}\label{eq:OVmAChPT}
\langle\mathcal{O}^{d}_{6,V-A}(\mu)\rangle \, &=\,  32\pi \alpha_{s}(\mu)\, F^{6}
a_{88}^{\lambda\lambda}(\mu) \left\{
1 - \frac{2 M_K^2}{(4\pi F)^2}  \, 
\log{\left(\frac{M_{K}^{2}}{\nu_{\chi}^2}\right)} - \frac{4 M_{\pi}^2}{(4\pi F)^2} \,\log{\left(\frac{M_{\pi}^{2}}{\nu_{\chi}^2}\right)}
\right.\nonumber\\ & \hskip 1cm
\left.
+ \,\frac{4}{F^2}\, M_\pi^2\; c_4^{\lambda\lambda}(\nu_{\chi},\mu)
+ \frac{2}{F^2}\, (2 M_K^2 +M_\pi^2)\; c_6^{\lambda\lambda}(\nu_{\chi},\mu)
\right\} 
 + \mathcal{O}(p^4,\alpha_{s}^{2})
\, .
\end{align}
The chiral logarithmic corrections are unambiguously predicted in terms of the LO coupling $a_{88}^{\lambda\lambda}(\mu)$, but there are in addition local contributions from the $\mathcal{O}(p^2)$ $\chi$PT
operators~\cite{Cirigliano:1999pv}
\begin{align}\nonumber
\mathcal{L}_{8_L,8_R}^{\mathcal{O}(p^4)}&=\frac{F^4}{4}\, c_{4}^{\lambda\lambda}\; [t_{LR}^{\lambda\lambda}]^{jl}_{ik}\;\lambda_{L,j}^{a,i}\lambda_{R,l}^{b,k}\;\mathrm{Tr}(\lambda_L^{a}S_{+} U^{\dagger}\lambda_R^{b}U+\lambda^{a}_L U^{\dagger}\lambda^{b}_R U S_{+})
\\&+\frac{F^4}{4}\, c_{6}^{\lambda\lambda}\; [t_{LR}^{\lambda\lambda}]^{jl}_{ik}\;\lambda_{L,j}^{a,i}\lambda_{R,l}^{b,k}\;\mathrm{Tr}(\lambda^{a}_L U^{\dagger}\lambda^{b}_RU)\;\mathrm{Tr}(S_{+}) \, ,
\end{align}
with $S_+ = U^\dagger\chi + \chi^\dagger U$. The renormalized couplings $c_{4,6}^{\lambda\lambda}(\nu_{\chi},\mu)$ reabsorb the loop divergences and, therefore, depend on both the short-distance ($\mu$) and $\chi$PT ($\nu_\chi$) renormalization scales:
\begin{equation}
c_{i}^{\lambda\lambda,(0)}(\mu)\, =\, 
a_{88}^{\lambda\lambda}(\mu) \left\{
c_{i}^{\lambda\lambda,r}(\nu_{\chi},\mu) + 
\frac{\zeta_i}{(4\pi F)^2}
\left[ \frac{2\,\nu_\chi^{D-4}}{D-4} + \gamma_E -\log{(4\pi)}-1\right]
\right\},
\end{equation}
where $\zeta_4 =\frac{3}{4}$ and $\zeta_6 =\frac{1}{2}$. These couplings can be easily estimated in the large-$N_C$ limit, using Eq.~(\ref{eq:LargeNcLR}):
\begin{equation}\label{eq:c4c6LargeNc}
 c_{4}^{\lambda\lambda,\infty}(\nu_{\chi},\mu) \, =\, 
 2\, (2 L_8 + H_2) \, =\, \frac{32}{3}\, L_8\, ,
 \qquad\qquad\quad
 c_{6}^{\lambda\lambda,\infty}(\nu_{\chi},\mu) \, =\, 0\, .
\end{equation}
The dependence of the product $a_{88}^{\lambda\lambda,\infty}(\mu) \, c_{4}^{\lambda\lambda,\infty}(\nu_{\chi},\mu)$ on the short-distance renormalization scale $\mu$ is fully carried by $a_{88}^{\lambda\lambda,\infty}(\mu)$, through the factor $B(\mu)$ in Eq.~(\ref{eq:Bmu}), while the dependence on the scale $\nu_\chi$ is of higher order in $1/N_C$ because it is a $\chi$PT loop effect.

The NLO corrections in $\alpha_s$ are also known \cite{Lanin:1986zs,Adam:1993uu}. For $\mathcal{O}^{d}_{6,V-A}$ they have the structure:
\begin{align}
[t_{LR}^{\delta\delta}]^{jl}_{ik}\, &=\, \alpha_s^2
\left[ A_{1} + B_{1}\, \log{\Big(\frac{-q^2}{\mu^2}\Big)}\right]\;
(\lambda_{L,i}^{1,j}\lambda_{R,k}^{1,l}+\lambda_{L,i}^{2,j}\lambda_{R,k}^{2,l}) \, ,
\\
[t_{LR}^{\lambda\lambda}]^{jl}_{ik}\, &=\, 8\pi\alpha_s \left[ 1+\frac{\alpha_{s}}{2\pi}\, A_{8}
+ \frac{\alpha_{s}}{2\pi}\, B_{8}\, \log{\Big(\frac{-q^2}{\mu^2}\Big)}\right]\;
(\lambda_{L,i}^{1,j}\lambda_{R,k}^{1,l} +\lambda_{L,i}^{2,j}\lambda_{R,k}^{2,l}) \, ,
\end{align}
where~\cite{Lanin:1986zs,Adam:1993uu,Boito:2015joa}
\begin{equation}
   B_1\, =\, 3 \left(1-\frac{1}{N_C^2}\right)\, ,
   \qquad\qquad\qquad
   B_8\, =\, \frac{n_f-N_C}{3} -\frac{3}{N_C}\, , 
\end{equation}
are related to the anomalous dimensions of the four-quark operators, with $n_f=3$ quark flavours. The values of the non-logarithmic coefficients $A_1$ and $A_8$ depend on the adopted regularization prescription for $\gamma_{5}$. The most recent calculation gives,
in the naive dimensional regularization (NDR) and 't Hooft-Veltman (HV) schemes~\cite{Cirigliano:2001qw}:
\begin{equation}
    A_1\, =\,\left\{ \begin{array}{cc} 2 & \text{(NDR)} \\[2pt] -10/3 & \text{(HV)}\end{array}\right.\, ,
    \qquad\qquad\qquad
    A_8\, =\,\left\{ \begin{array}{cc} 25/4 & \text{(NDR)} \\[2pt] 21/4 & \text{(HV)}\end{array}\right.\, ,
\end{equation}
for $n_f=N_C=3$.
These NLO QCD corrections introduce the colour-singlet four-quark left-right operator and, therefore, additional non-perturbative parameters.
The final expression for the vacuum condensate at NLO in $\chi$PT and $\alpha_{s}$ is then given by
\begin{align}
\langle \mathcal{O}^{d}_{6,V-A}(\mu)\rangle &=\, 32\pi \alpha_{s}(\mu)\, F^6
\; a_{88}^{\lambda\lambda}(\mu)
\left[ 1+\frac{\alpha_{s}(\mu)}{2\pi}A_{8}+\frac{\alpha_{s}(\mu)}{2\pi}B_{8}\,\log{\left(\frac{-q^2}{\mu^2}\right)}\right]
\nonumber\\ 
&\hskip 1.2cm\cdot\, 
\left\{
1- \frac{2 M_{K}^{2}}{(4\pi F)^{2}}\,\log{\left(\frac{M_{K}^{2}}{\nu_{\chi}^{2}}\right)}- \frac{4 M_{\pi}^{2}}{(4\pi F)^{2}}\,\log{\left(\frac{M_{\pi}^{2}}{\nu_{\chi}^{2}}\right)} 
\right.\nonumber\\ &
\hskip 1.5cm\left.
+ \,\frac{4}{F^2}\, M_\pi^2\; c_4^{\lambda\lambda}(\nu_{\chi},\mu)
+ \frac{2}{F^2}\, (2 M_K^2 +M_\pi^2)\; c_6^{\lambda\lambda}(\nu_{\chi},\mu)
\right\} 
\nonumber\\ 
& +\, 4\alpha_{s}^2(\mu)\, F^6 
\; a_{88}^{\delta\delta}(\mu)
\left[A_{1}+B_{1}\log{\left(\frac{-q^2}{\mu^2}\right)}\right]
\nonumber\\ 
&\hskip 1.2cm\cdot\,  
\left\{1-\frac{2 M_{K}^{2}}{(4\pi F)^{2}} \,\log{\left(\frac{M_{K}^{2}}{\nu_{\chi}^{2}}\right)}- \frac{4 M_{\pi}^{2}}{(4\pi F)^{2}}\,\log{\left(\frac{M_{\pi}^{2}}{\nu_{\chi}^{2}}\right)} 
\right.\nonumber\\ &\hskip 1.5cm\left.
+ \,\frac{4}{F^2}\, M_\pi^2\; c_4^{\delta\delta}(\nu_{\chi},\mu)
+ \frac{2}{F^2}\, (2 M_K^2 +M_\pi^2)\; c_6^{\delta\delta}(\nu_{\chi},\mu)
\right\} .
\label{eq:fullcondapp}
\end{align}
The contribution from the colour-singlet four-quark operator is nevertheless very small. In addition to be a higher-order correction in the strong coupling, it is colour suppressed. In the large-$N_C$ limit,
\begin{equation}
  a_{88}^{\delta\delta,\infty}(\mu) \, =\,  c_4^{\delta\delta,\infty}(\nu_{\chi},\mu)
  \, =\,  c_6^{\delta\delta,\infty}(\nu_{\chi},\mu)\, =\, 0\, .
\end{equation}
In order to keep track of the total size of the chiral logarithmic corrections, which will be useful to estimate uncertainties in the comparison with the kaon sector in Section \ref{sec:latt}, it is convenient to rewrite Eq. (\ref{eq:fullcondapp})
reabsorbing the chiral logarithms into powers of $F/F_{\pi}$. Doing that and approximating the NLO counterterms, which play a very minor numerical role,  by their large-$N_{c}$ values, one finds:
\begin{align}
\langle \mathcal{O}^{d}_{6,V-A}(\mu)\rangle &=\, 32\pi \alpha_{s}(\mu)\, F_{\pi}^{4} 
\left\{ F^2  a_{88}^{\lambda\lambda}(\mu)
\left[ 1+\frac{\alpha_{s}(\mu)}{2\pi}A_{8}+\frac{\alpha_{s}(\mu)}{2\pi}B_{8}\,\log{\left(\frac{-q^2}{\mu^2}\right)}\right]
\right.\nonumber\\ 
&\left.
+ \, F^2 a_{88}^{\delta\delta}(\mu)\; \frac{\alpha_{s}(\mu)}{8\pi}
\left[A_{1}+B_{1}\log{\left(\frac{-q^2}{\mu^2}\right)}\right]\right\}
\left\{  1-\frac{16 M_{\pi}^2}{F_{\pi}^2}\, \left(L_5-\frac{8}{3}\, L_8 \right) 
\right\} 
\, .
\label{eq:fullcond}
\end{align}

\subsection{Other flavour-breaking structures}

The bosonization of $\mathcal{O}^{d-s}_{6,V+A}$ can be obtained with the same method. However, the $(8_L,8_R)$ structures disappear when summing the vector and axial-vector contributions, as can be seen in Eq.~(\ref{eq:tildet-d-LR}). This implies that the corresponding effective operator contains two derivatives and, therefore, cannot acquire a vacuum expectation value at tree-level. The associated $\mathcal{O}_{6,V+A}^{d-s}$ condensate can be only generated through $\chi$PT loops and is then heavily suppressed with respect to $\mathcal{O}_{6,V-A}^{d}$ by a factor of $\mathcal{O}(M_{K}^4/\Lambda_{\chi}^4)$.

The bosonization of $\mathcal{O}_{6,EM}-\mathcal{O}_{6,V}^{d}$ contains an $\mathcal{O}(p^0)$ term proportional to $a_{88}^{\lambda\lambda}$, generated by the $[t_{LR}^{\lambda\lambda}]$ contribution. However, the vacuum expectation value of this term also vanishes at tree-level and, as a consequence, it has a chiral 
suppression of $\mathcal{O}(M_{K}^{2}/\Lambda_{\chi}^2)$.
This suppression is not accidental. The currents $\bar{J}_{EM}^{\mu}$ and $J_{ud,V}^{\mu}$ are trivially related by an $SU(3)_{V}$ rotation. Their two-point correlation functions $\overline{\Pi}_{EM}$ and $\Pi^{d}_{V}$ must then be identical, as far as the $SU(3)_{V}$ symmetry is preserved. But $SU(3)_{V}$ cannot be spontaneously broken in QCD \cite{Vafa:1983tf}. Any nonzero condensate in the difference must then emerge as a consequence of an explicit symmetry breaking of $SU(3)_{V}$, which is fully dominated by the nonzero strange quark mass, leading to an $\mathcal{O}(M_{K}^{2}/\Lambda_{\chi}^2)$ suppression.

\section{\boldmath Determination of \texorpdfstring{$\mathrm{Im}(g_{8}\, g_{\mathbf{ewk}})$}{Img8gew} from \texorpdfstring{$\tau$}{tau}-decay data}
\label{sec:pheno}

The inclusive invariant-mass distributions of the final hadrons in $\tau$ decay directly measure the hadronic spectral functions associated with the $ud$ and $us$ two-point current correlators in Eq.~(\ref{eq:Correlators-def}), up to the $\tau$ mass scale \cite{Braaten:1991qm,Pich:2020gzz}:
\begin{equation}
 \frac{d\Gamma}{ds}\, =\, \frac{G_{F}^{2}}{16\pi^2}\, m_{\tau}^{3}\, S_{\mathrm{EW}} 
 \left(1-\frac{s}{m_{\tau}^{2}}\right)^{2} \left\{
\left(1+2\,\frac{s}{m_{\tau}^{2}}\right)\, \mathrm{Im}\,\Pi_{\tau}^{L+T}(s)-2\,\frac{s}{m_{\tau}^{2}}\;\mathrm{Im}\,\Pi_{\tau}^{L}(s)     \right\} \, ,
\end{equation}
where
\begin{equation}
\Pi_\tau(s)\,\equiv\,   \sum_{i=d,s}  |V_{ui}|^{2}\,\left[ \Pi_{ui,V}(s) + \Pi_{ui,A}(s)\right] 
\end{equation}
and $S_{\mathrm{EW}} = 1.0201\pm 0.003$ incorporates the (renormalization-group improved) electroweak corrections \cite{Marciano:1988vm,Braaten:1990ef,Erler:2002mv}. Identifying an even or odd number of pions and kaons in the final state, one can further separate the spectral distributions corresponding to $V_{ud}$, $A_{ud}$ and $V_{us}+A_{us}$.

\begin{figure}[tb]\centering
\includegraphics[width=1\textwidth]{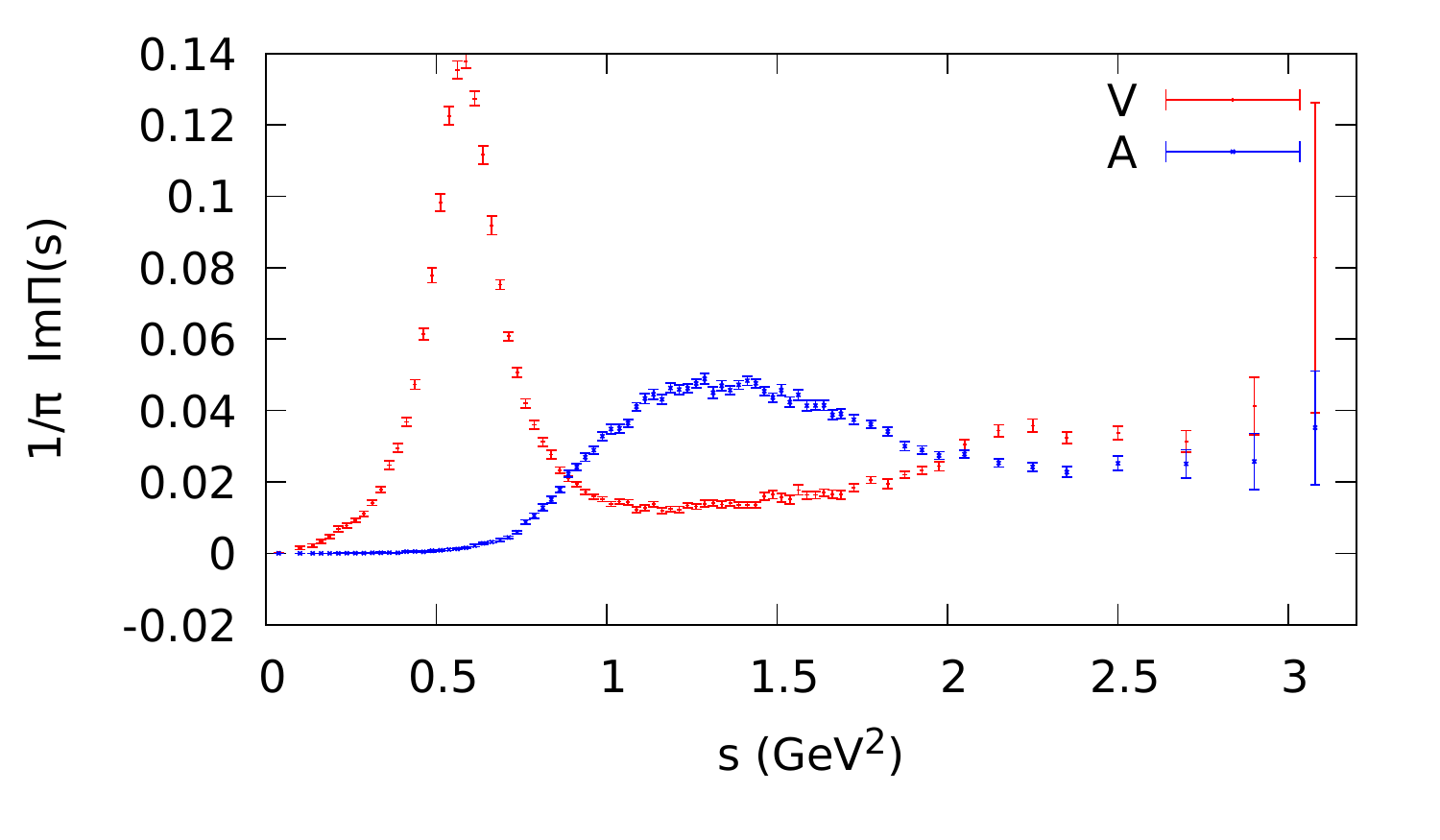}
\caption{ALEPH non-strange spectral functions $\frac{1}{\pi}\,\mathrm{Im}\,\Pi_{ud,V/A}^{L+T}(s)$.}
\label{fig:specs}
\end{figure}

We are going to focus in the Cabibbo-allowed $ud$ spectral functions, making use of the most precise measurements of the corresponding vector and axial-vector distributions, extracted from ALEPH data \cite{Davier:2013sfa}, which are displayed in Fig.~\ref{fig:specs}. Given the current experimental uncertainties, the longitudinal axial spectral function is well approximated by the pion pole contribution, $\mathrm{Im}\Pi_{A}^{L}(s)=2\pi F_{\pi}^2 \, \delta(s-m_{\pi}^{2})$, while the tiny contribution from $\mathrm{Im}\Pi_{V}^{L}(s)$ can be safely neglected.

\begin{figure}[tb]\centering
\includegraphics[width=0.48\textwidth]{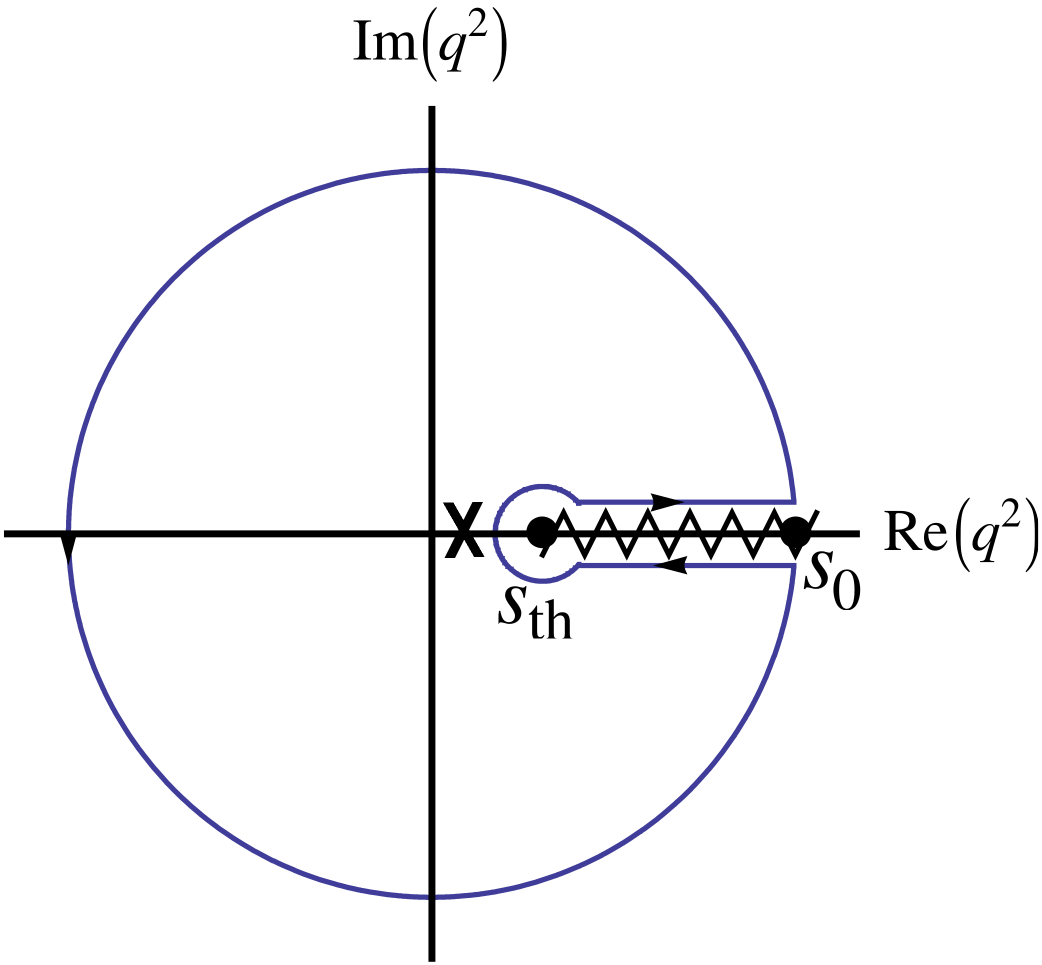}
\caption{Analytic structure of $\Pi^{L+T}(s)$ and complex contour used to derive Eq.~(\ref{eq:sumrulepre}).}
\label{fig:circuit1}
\end{figure}

The current correlators are analytic functions in all the complex $s\equiv q^2$ plane, except for the physical cut in
the positive real axis where they acquire their absorptive components. Apart from the pion pole, this cut starts at $s_{\mathrm{th}}=4 M_{\pi}^{2}$. Integrating along the circuit of Fig.~\ref{fig:circuit1} a given correlator times any arbitrary weight function $\omega(s)$, analytic at least in the same complex region as the correlator, one finds
\begin{equation}\label{eq:sumrulepre}
\int^{s_{0}}_{s_{\mathrm{th}}}\frac{ds}{s_{0}}\; \omega(s)\,\frac{1}{\pi}\operatorname{Im}\Pi^{L+T}(s)
+\frac{1}{2\pi i}\oint_{|s|=s_{0}}\! \frac{ds}{s_{0}}\; \omega(s) \,\Pi^{L+T}(s)
\; =\; 2\, \frac{F_{\pi}^{2}}{s_{0}}\,\omega(M_{\pi}^{2}) \, .
\end{equation}
In the first term one can introduce the experimental spectral function, while for large enough values of $s_{0}$,
the OPE of $\Pi^{L+T}(s)$ becomes an excellent approximation for the integral along the complex circle $|s|=s_0$, except maybe for the region near the positive real axis \cite{Braaten:1991qm,LeDiberder:1992zhd}. 

The small differences between using the physical correlators or their OPE approximations are known as quark-hadron duality violations \cite{Dubovikov:1983rj,Chibisov:1996wf,Shifman:2000jv,Cata:2005zj,Cirigliano:2002jy,GonzalezAlonso:2010rn,GonzalezAlonso:2010xf,Boito:2017cnp}:
\begin{align}\label{eq:DV}\nonumber
\delta_{\mathrm{DV}}[\omega(s),s_{0}]\;&\equiv\; \frac{1}{2\pi i} \oint_{|s|=s_{0}}\frac{ds}{s_{0}}\; \omega(s)\, \left[\Pi^{L+T}(s)-\Pi^{L+T}_\mathrm{OPE}(s)\right]
\; \\&=\frac{1}{\pi}\;\int^{\infty}_{s_{0}}\frac{ds}{s_{0}}\;\omega(s)\,\left[\mathrm{Im}\,\Pi^{L+T}(s)-\mathrm{Im}\,\Pi^{L+T}_{\mathrm{OPE}}(s)\right]\, .
\end{align}
These effects get strongly suppressed when using (pinched) weight functions $\omega(s)$ with zeros at $s=s_{0}$. This can be seen in two different ways. First, the zeros at $s=s_{0}$ kill the contributions to the contour integral from the region near the physical axis, where the OPE is less justified. Second, since $\mathrm{Im}\,\Pi^{L+T}_{\mathrm{OPE}}(s)$ approaches $\mathrm{Im}\,\Pi^{L+T}(s)$ very fast, typically exponentially, the spectral differences are dominated by the region near $s_{0}$ that pinched weight functions remove.

In this work, we are interested in the correlation function $\Pi_{V-A}(s)\equiv \Pi^{L+T}_{ud,V}-\Pi^{L+T}_{ud,A}$, which vanishes to all orders in perturbation theory when quark masses are neglected. Since $m_{u,d}$ are tiny, this is an excellent approximation in the up-down sector. The non-zero value of $\Pi(s)$ originates in the spontaneous breaking of chiral symmetry by the QCD vacuum, which results in different vector and axial-vector correlators. The leading OPE contribution comes from four-quark operators with $D=6$ (the lowest dimension where a chiral-symmetry breaking can be induced with massless quark and gluon fields) and is suppressed by six powers of the $\tau$ mass. Although the vector and axial-vector spectra in Fig.~\ref{fig:specs} have very different shapes in the low-energy resonance regime, chiral symmetry implies a very strong suppression of their integrated difference in Eq.~(\ref{eq:sumrulepre}) when duality violations are suppressed, {\it i.e.}, taking $s_{0}$ near $m_{\tau}^{2}$ and pinched weight functions. 

\begin{figure}[tb]\centering
\includegraphics[width=1.\textwidth]{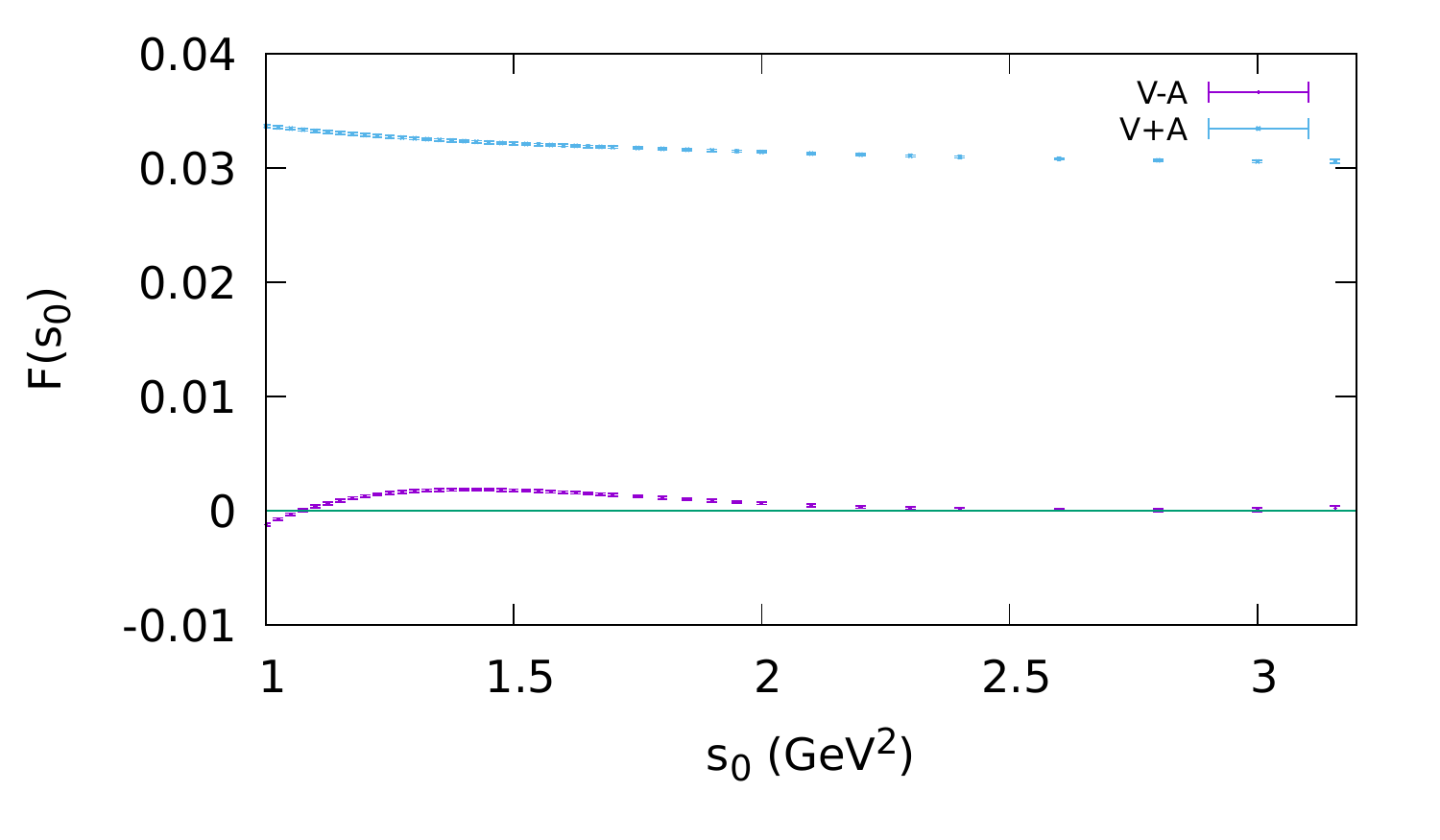}
\caption{$F_{V\pm A}(s_{0})$ as defined in Eq.~(\ref{eq:WSR}). The upper and lower curves correspond to $V+A$ and $V-A$, respectively.}
\label{fig:wsrmixed}
\end{figure}

In order to illustrate this, let us focus on the pinched integrals
\begin{equation}
F_{V\pm A}(s_{0})\,\equiv\, \int^{s_{0}}_{s_{\mathrm{th}}}\frac{ds}{s_{0}}\; \left(1-\frac{s}{s_{0}} \right)\frac{1}{\pi}\operatorname{Im}\Pi_{V\pm A}(s)
\,\pm\, 2\frac{F_{\pi}^{2}}{s_{0}}\left(1-\frac{m_{\pi}^{2}}{s_{0}}\right)  ,
\label{eq:WSR}
\end{equation}
which are plotted in Fig. \ref{fig:wsrmixed}, as a function of the upper integration limit $s_0$.
%
%
In spite of the very small experimental uncertainties, which are below the percent level, no signatures of non-perturbative effects can be observed near the $\tau$ mass. $F_{V-A}\approx 0$, as expected, exhibiting the negligible role of duality violations in this pinched observable, at large $s_0$. While for the $V+A$ channel this fact leads to a precise determination of the strong coupling \cite{Braaten:1991qm,Davier:2013sfa,Boito:2014sta,Pich:2016bdg}, it also translates into a very limited sensitivity to the gluon and four-quark condensates. Since four-quark operators only enter into the integral (\ref{eq:WSR}) through the $\alpha_{s}$-suppressed logarithmic term in Eq.~(\ref{eq:wilsons}), they could only leave a sizeable imprint at very small values of $s_{0}$, where duality violations are beyond theoretical control, as becomes manifest looking at the large splitting of the spectra in Fig.~\ref{fig:specs}.

Taking into account the current experimental uncertainties and provided the renormalization scale $\mu$ is set close to the tau mass, one can neglect the running of the QCD corrections and work with  effective condensates $\langle\mathcal{O}_{D}\rangle$ in Eq.~(\ref{eq:wilsons}), independent of the energy scale.\footnote{
The only large NLO correction in $\alpha_{s}$, arising from the large value of $A_{8}$ in eq. (\ref{eq:fullcond}), does not change this approximation and will be taken into account.}
Inserting the OPE into the second term of Eq.~(\ref{eq:sumrulepre}) and invoking the Cauchy formula, every monomial weight function $\omega(s)= (s/s_{0})^{n}$ becomes connected to a different effective condensate $\langle\mathcal{O}_{2(n+1)}\rangle$. For the $V-A$ correlation function, one finds
\begin{align}
\int^{s_{0}}_{s_{\mathrm{th}}}\frac{ds}{s_{0}}\; \left(\frac{s}{s_{0}}\right)^{n}\,\frac{1}{\pi}\operatorname{Im}\Pi_{V-A}(s)
&+(-1)^{n+1}\frac{\mathcal{O}_{2(n+1)}}{s_{0}^{n+1}} 
+ \delta_{\mathrm{DV}}[(s/s_0)^n,s_{0}]
\; =\; 2\, \frac{F_{\pi}^{2}}{s_{0}}\,\left(\frac{M_{\pi}^{2}}{s_{0}}\right)^{n} .\label{sumrule22}
\end{align}
Determining $\langle\mathcal{O}_{6}\rangle$, which is nothing else but $\langle \mathcal{O}^d_{6,V-A}(s_0)\rangle$ in Eq.~(\ref{eq:fullcond}),\footnote{The scale is set to $s_0$ in order to avoid large logarithms in the $\alpha_{s}$ corrections.}  is going to give us $a_{88}^{\lambda\lambda}(s_0)$, which is linked to $e^{2}g_{8}\, g_{\mathrm{ewk}}(s_0)$ through Eq.~(\ref{eq:g8gew}).

\subsection{\boldmath Determination of \texorpdfstring{$\langle\mathcal{O}_{6}\rangle$}{O6}}

We already determined $\langle\mathcal{O}_{6}\rangle$ in Ref. \cite{Rodriguez-Sanchez:2016jvw}, which updated Refs. \cite{GonzalezAlonso:2010rn,GonzalezAlonso:2010xf}. In this subsection we revisit it, introducing some minor modifications and extra tests. 

\subsubsection{\boldmath Determination of \texorpdfstring{$\langle\mathcal{O}_{6}\rangle$}{O6} based on energy stability}

Naively, one could try to estimate $\langle\mathcal{O}_{6}\rangle$ by using Eq.~(\ref{sumrule22})
with the corresponding monomial function $\omega(s)= (s/s_{0})^2$, 
hoping that at large-enough energies duality violations are negligible.
This should be reflected in the appearance of a plateau at high energies, when making the trivial rescaling of that equation, so that it converges to $\langle\mathcal{O}_{6}\rangle$ for large-enough values of $s_{0}$.
\begin{figure}[tb]\centerline{
\includegraphics[width=1\textwidth]{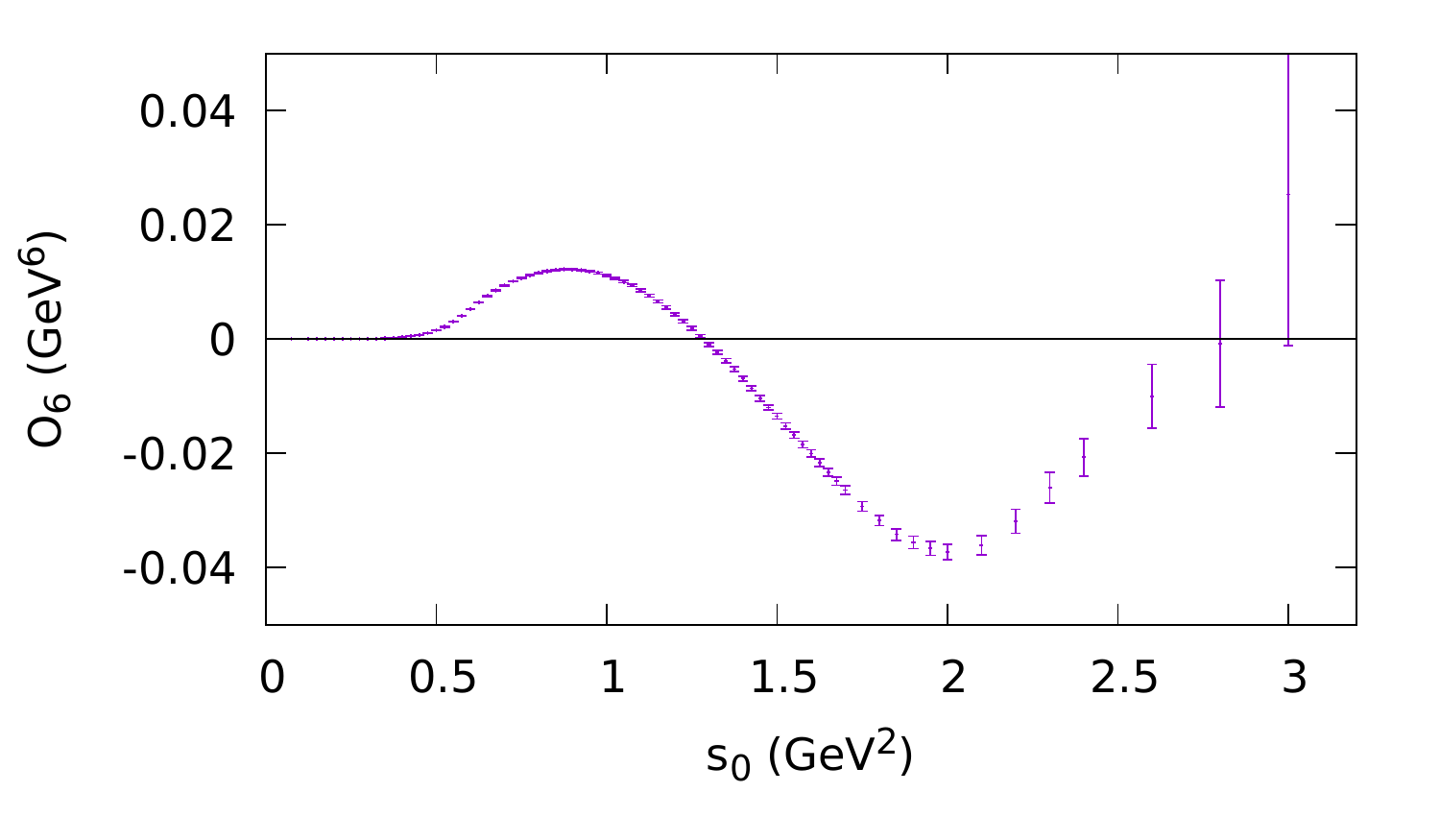}}
\caption{\label{fig:o6nopinch} Rescaled version of the moment associated to $\omega(s)= (s/s_{0})^{2}$ so that, at large-enough $s_{0}$,  it converges to $\langle\mathcal{O}_{6}\rangle$.}
\end{figure}
However, as can be seen in Fig. \ref{fig:o6nopinch}, 
there are large violations of quark--hadron duality and the experimental uncertainties grow when increasing $s_0$.
The weight function is enhancing both the contribution of the high-energy part of the spectral function, where data are less precise, and the high-energy duality-violation tail associated to Eq.~(\ref{eq:DV}).

\begin{figure}[tb]\centerline{
\includegraphics[width=1.\textwidth]{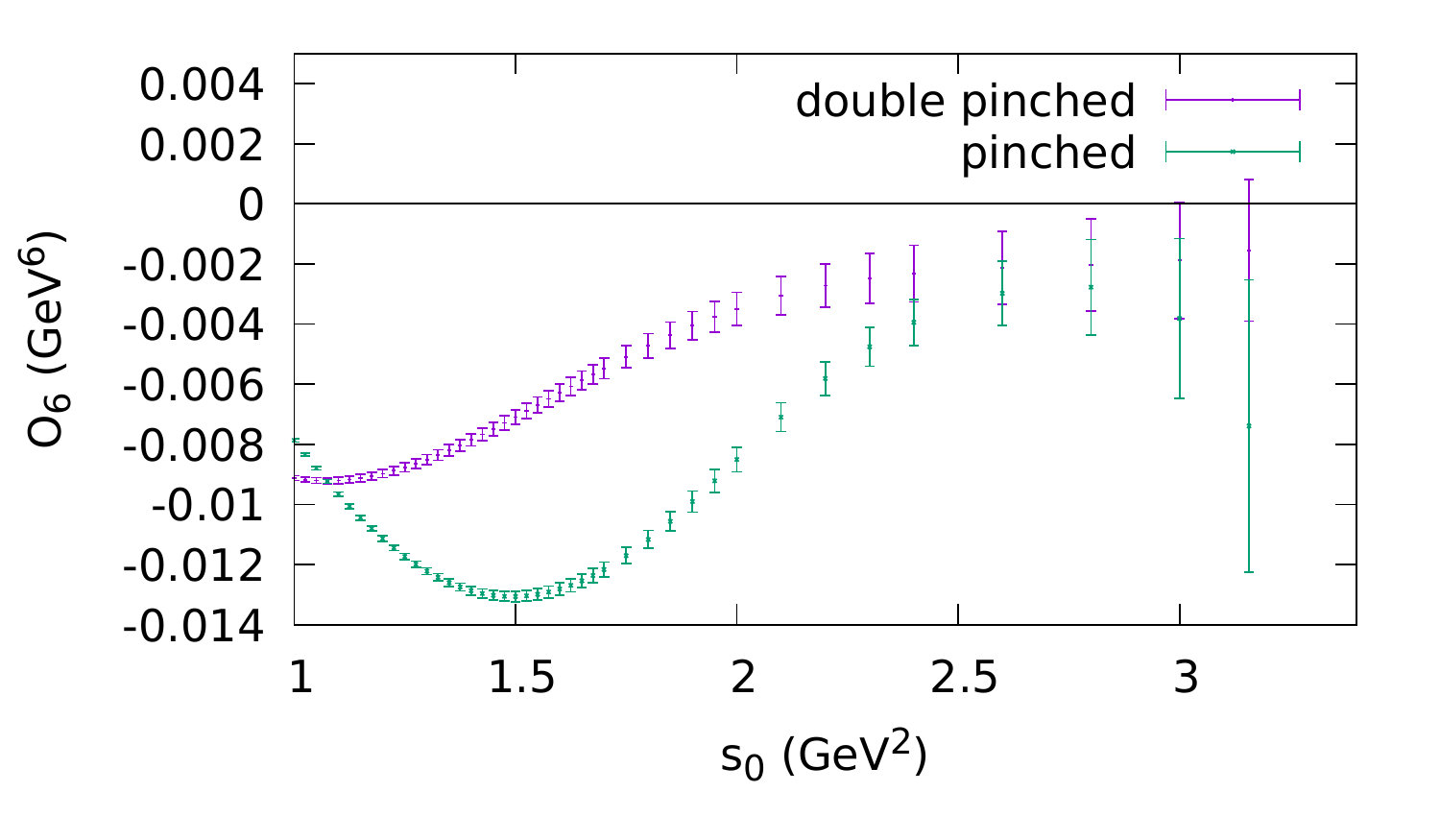}}
\caption{\label{fig:o6pinch} Results for $\langle\mathcal{O}_{6}\rangle$, as a function of $s_{0}$, obtained from (appropriately rescaled) moments with pinched weights.}
\end{figure}

As already mentioned before, duality violations can be suppressed introducing pinched weight functions, containing the desired monomia. Taking the (once-pinched) $\omega(s) = x (1-x)$
and (double-pinched) $\omega(s) = (1-x)^2$
weight functions, with $x=s/s_0$,
we obtain the values of $\langle\mathcal{O}_{6}\rangle$ shown in Fig. \ref{fig:o6pinch}, to be compared with Fig. \ref{fig:o6nopinch} (notice the different scales in the $y$ axes). Experimental uncertainties are clearly reduced and a plateau has arisen. One may still argue, by taking an artificial shape for the high-energy tail of the spectral function, that the plateau could be accidental and disappear at higher values of $s_0$.
However, since there is an increasing hadronic multiplicity at $s_{0}\sim m_{\tau}^{2}$, duality violations should go to zero very fast when increasing the energy, making this contrived scenario very unlikely. Moreover, the results from the two pinched weight functions approach the same value of $\langle\mathcal{O}_{6}\rangle$ at large $s_0$. Thus, duality violations become indeed relatively small at large $s_{0}$, specially for the doubly-pinched weight that leads to smaller uncertainties.
Taking that into account, we take as central value the lowest energy point within the plateau, {\it i.e.}, the lowest one which lies within the experimental error bars of the following ones ($s_{0}=2.1\; \mathrm{GeV}^{2}$), and as an estimate of duality-violation uncertainties its difference with the last energy point with an acceptable experimental resolution, {\it i.e.}, $s_{0}=2.8\; \mathrm{GeV}^{2}$. We obtain in this way
\begin{equation}
\langle\mathcal{O}_{6}\rangle^{\mathrm{stab}}\, =\, (-3.1 \pm 0.6_{\,\text{exp}} \pm 1.0_{\text{\,DV}})\cdot 10^{-3} \; \mathrm{GeV}^{6}
\, =\, (-3.1 \pm 1.2)\cdot 10^{-3} \; \mathrm{GeV}^{6}  \, .
\end{equation}

\subsubsection{\boldmath Determination of \texorpdfstring{$\langle\mathcal{O}_{6}\rangle$}{O6} modeling duality violations}

An alternative approach to estimate duality-violation effects consists in trying to guess the spectral function 
$\rho(s) =\frac{1}{\pi}\,\mathrm{Im}\,\Pi_{V-A}(s)$ above the region where data are available.\footnote{At LO in $\alpha_{s}$, $\rho_{DV}(s) = \rho(s)$ because the OPE does not generate absorptive contributions.} 
In order to do that, a parametrization is unavoidable and, therefore, some model-dependence arises.
We will impose the theoretical requirement that the physical spectral function must obey the Weinberg Sum Rules (WSRs) \cite{Weinberg:1967kj}, {\it i.e.} Eq.~(\ref{sumrule22}) for $n=0$ and $n=1$, which do not involve any condensate contribution. This  condition restricts very 
strongly the possible choice of admissible spectral functions.

We will adopt the four-parameter ansatz  \cite{Blok:1997hs,Shifman:1998rb,Shifman:2000jv,Cata:2005zj,Cata:2008ye,GonzalezAlonso:2010rn,GonzalezAlonso:2010xf}
\begin{equation}\label{eq:ansatz}
\rho(s)\, =\, \frac{1}{\pi}\,\kappa\, e^{-\gamma s}\,\sin{[\beta (s-s_{z})]}\qquad\qquad (s>\hat{s}_{0}) 
\end{equation}
that combines an oscillatory function with the expected exponential suppression at large values of $s$.

Following the procedure of Ref.~\cite{Rodriguez-Sanchez:2016jvw},
we generate $10^9$ random tuples of $(\kappa,\gamma,\beta,s_{z})$ parameters, so that every one of them represents a possible spectral function above a threshold $\hat{s}_{0}$. The fit to the ALEPH data does not show significant deviations (p-value above $5 \%$) from this specific ansatz above $\hat{s}_{0}=1.25\, \mathrm{GeV}^{2}$. However, the model is only motivated as an approximation at higher energies, where the hadronic multiplicity is also higher. 
As in Ref~\cite{Rodriguez-Sanchez:2016jvw}, we only accept those tuples contained within the $90 \%$ C.L. region ($\chi^{2}<\chi^{2}_{\text{min}}+7.78$) in the fit to the experimental data. By doing that, we are relaxing somewhat the model dependence by allowing small deviations of the admissible spectral functions from the fitted data.

In Ref.~\cite{Rodriguez-Sanchez:2016jvw} we imposed in this step  the short-distance constraints on the tuples, {\it i.e.}, the WSRs. However, the experimental uncertainties on these constraints become then correlated in a non-trivial way with the experimental uncertainty of the final parameters. 
 
\begin{figure}[tb]\centerline{
\includegraphics[width=1.\textwidth]{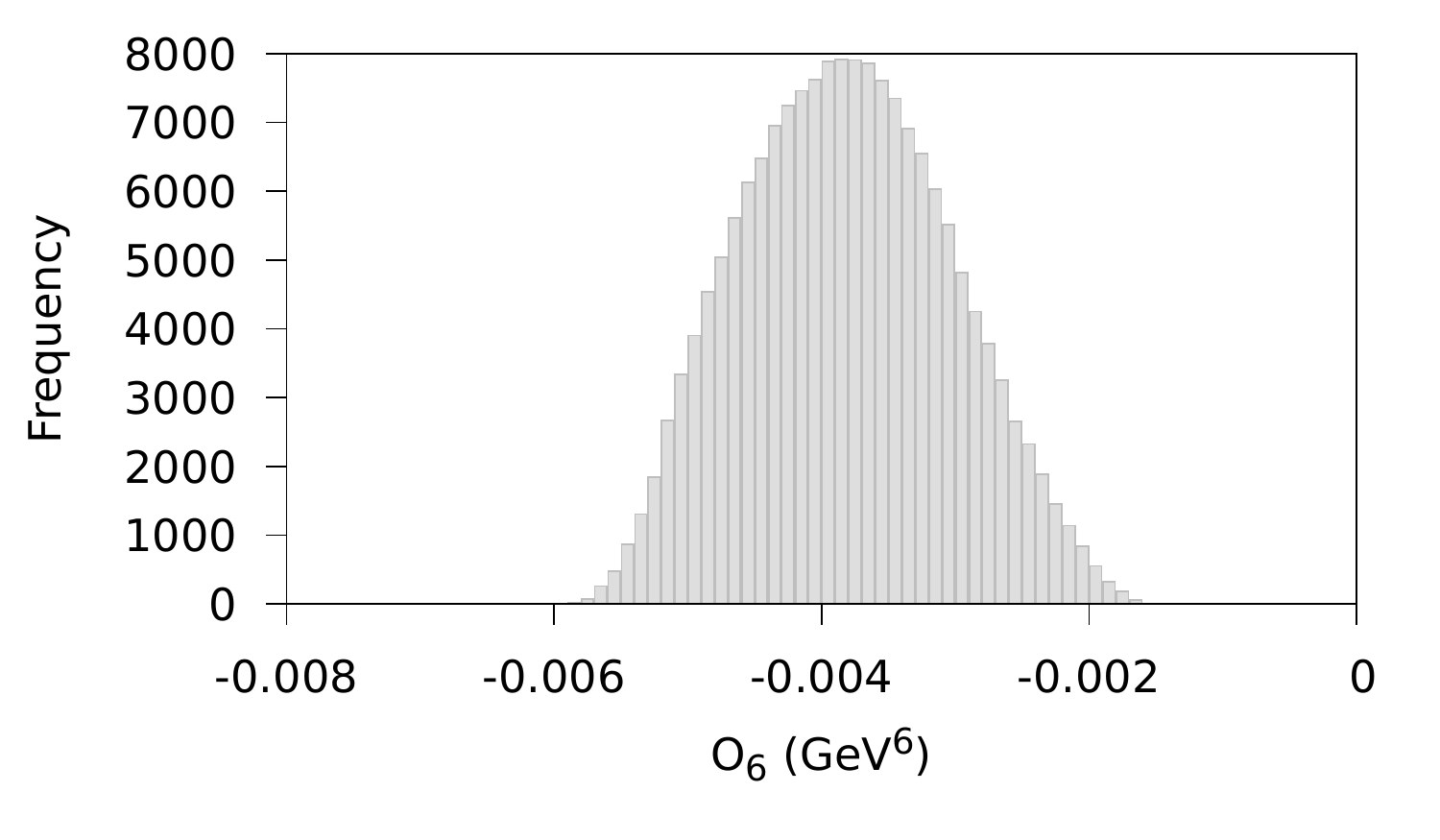}}
\caption{\label{fig:disto6} Distribution for $\langle\mathcal{O}_{6}\rangle$ obtained with the tuples procedure at $\hat{s}_{0}=1.7 \, \mathrm{GeV}^{2}$.}
\end{figure}

In order to avoid that, for every accepted spectral function,  we perform a combined fit to Eq.~(\ref{sumrule22}) for $n=0,1,2$ to extract $\langle\mathcal{O}_{6}\rangle$. Then we only accept those spectral functions that are compatible with the WSRs ($n=0,1$), selecting only the ones whose p-values in the combined fit are larger than $5 \%$. Every accepted spectral function gives a value of $\langle\mathcal{O}_{6}\rangle$. Fig.~\ref{fig:disto6} shows the statistical distribution of $\langle\mathcal{O}_{6}\rangle$ values, obtained with $\hat{s}_{0}=1.7 \; \mathrm{GeV}^{2}$. The width of this distribution provides a good assessment of the duality-violation uncertainty.

The choice of $\hat{s}_{0}$, the parameter separating the use in Eq.~(\ref{sumrule22}) of real data or the model ansatz,
is somehow arbitrary. Therefore, a smooth dependence on the chosen value of $\hat{s}_{0}$,  within a large-enough range, is a minimal requirement that we should impose.\footnote{This is analogous to the stability condition (plateau) imposed in the previous subsection. Using the ansatz at lower $\hat s_0$ does not improve the results if the convergence of data to the model at that energy is actually worse than the convergence of data to the QCD OPE above $\hat s_0$ \cite{Pich:2016bdg}.}
Repeating our procedure with different thresholds leads to the results displayed in Table \ref{tab:th}.
The overall agreement is acceptable. We choose $\hat{s}_{0}= 1.7 \; \mathrm{GeV}^{2}$ as our optimal threshold, large enough to have some hadronic multiplicity and small enough to be able to constrain the space of parameters. We then obtain:
\begin{equation}\label{eq:O6detDV}
\langle\mathcal{O}_{6}\rangle^{\mathrm{ans}}\, =\, (-3.8 ^{\,+0.8}_{\,-0.9\;\text{DV}}\pm 0.1_{\,\text{exp}}) \cdot 10^{-3}\, \mathrm{GeV}^{6}  \, .
\end{equation}

\begin{table}[tb]
\renewcommand\arraystretch{1.4}
\centering
\begin{tabular}{|c|c|c|c|c|c|}
\hline
$\hat{s}_{0}\;\, (\mathrm{GeV}^{2})$ & $1.25$ & $1.4$ &  $1.55$  &  $1.7$ &  $1.9$\\ \hline
$\langle\mathcal{O}_{6}\rangle\;\, (10^{-3}\;\mathrm{GeV}^{6})$ & $-5.2 \, {}^{+0.5}_{-0.3}$ & $-5.0 \, {}^{+0.5}_{-0.5}$ &$-5.3 \, {}^{+0.5}_{-0.3}$ &$-3.8 \, {}^{+0.8}_{-0.9}$ &$-3.1 \, {}^{+1.0}_{-1.2} $\\ \hline
\end{tabular}\caption{\label{tab:th}Results for $\langle\mathcal{O}_{6}\rangle$, obtained with our tuple procedure with different values of $\hat{s}_{0}$.}
\end{table}

However, even 
if the ansatz (\ref{eq:ansatz}) were exactly true above some threshold $\hat{s}_{0}$, 
this $\hat{s}_{0}$ could happen to be larger than the available energy range, so that the physical spectral function could not be well approximated by the fitted parameters. In that case, assuming small duality violations with double-pinch weights could be giving more accurate results than assuming the spectral function ansatz with the fitted parameters. This motivates averaging the two results. Fortunately, in this case both methods are in good agreement. Our final value, taking conservatively the quadratic sum of the lowest uncertainty plus half of the difference between central values, is:
\begin{equation}\label{a6val}
\langle\mathcal{O}_{6,V-A}^d(m_\tau^2)\rangle\,= \, (-3.5 \pm 0.9 ) \cdot 10^{-3} \;\mathrm{GeV}^{6} \, ,
\end{equation}
in total agreement with our previous determination in Ref.~\cite{Rodriguez-Sanchez:2016jvw} and the result obtained in Ref.~\cite{Boito:2015fra} with a different procedure.

\subsection{\boldmath Determination of \texorpdfstring{$g_{8}\,g_{\mathbf{ewk}}$}{g8gew}}

Inserting in Eq.~(\ref{eq:OVmAChPT}) 
the obtained value for $\langle\mathcal{O}_{6,V-A}^d\rangle$, 
we can perform a determination of $a_{88}^{\lambda\lambda}$ at NLO in the chiral counting. 
We approximate the tiny counterterm piece, which has a minor numerical role, by its large-$N_C$ estimate in  Eq.~(\ref{eq:c4c6LargeNc}). 
Incorporating also the large and dominant NLO correction in $\alpha_{s}$ coming from $A_{8}$ in Eq. (\ref{eq:fullcondapp}), which does not modify the energy-independent condensate approximation used in our determination of $\langle\mathcal{O}_{6,V-A}^d\rangle$, one finds
\begin{equation}\label{eq:a88det}
F^{2}\, a_{88}^{\lambda\lambda}(m_{\tau})\, =\, (-1.15 \pm 0.30_{\,\langle\mathcal{O}_{6}\rangle}\pm 0.11_{\,\mathrm{pert}})  \, \mathrm{GeV}^{2} \, ,
\end{equation}
where we have assigned an extra $10 \%$ perturbative uncertainty based on the expected size $\sim \frac{\alpha_{s}(m_{\tau}^{2})}{\pi}$ of the unaccounted NLO corrections.
Notice that more precise experimental data may allow in the future for a full NLO analysis.

Taking into account that $\mathrm{Im} (C_{7})$ is smaller than $\mathrm{Im} (C_{8})$, the large $N_{C}$-suppression of $a_{88}^{\delta\delta}$ with respect to $a_{88}^{\lambda\lambda}$ and the extra $\frac{1}{N_{C}}$ prefactor in the contribution proportional to $C_{8}\, a_{88}^{\delta\delta}$, we can safely neglect  the 
$a_{88}^{\delta\delta}$ term in Eq.~(\ref{eq:g8gew}) to derive\footnote{
Since we have neglected long-distance electromagnetic contributions, no reliable estimate of the real part can be made at this point.
While in general this is a good approximation due to the large enhancement of the short-distance piece with respect to the long-distance one, {\it i.e.}, $\alpha\log{(M_{W}^{2}/\mu^{2})}$ vs $\alpha\log{(\mu^{2}/M_{\rho/K}^{2})}$, no such logarithmic enhancement is present in $z_{8}(\mu)$, since the GIM mechanism sets $z_{8}(\mu >m_{c})=0$~\cite{Cirigliano:2003gt}.
}
%
\begin{equation}
e^{2}\,\mathrm{Im}\, (g_{8}\,g_{\mathrm{ewk}})\,\approx\, 12\, \mathrm{Im}\, [C_{8}(m_{\tau})]\; a_{8_L ,8_R}^{\lambda\lambda} (m_{\tau}) \, ,
\end{equation}
from which we find
\begin{equation}
\frac{e^{2}\,\mathrm{Im}\, (g_{8}\,g_{\mathrm{ewk}})}{\mathrm{Im}\tau}\; F^{2}\, =\, -  (1.07\, \pm\, 0.30) \cdot 10^{-2} \,  \mathrm{GeV}^{2} \, .
\end{equation}
This phenomenological determination has a smaller central value than previous estimates, but, within the quoted uncertainties, it is in agreement with most of them  \cite{Bijnens:2000im,Bijnens:2001ps,Knecht:2001bc,Narison:2000ys,Cirigliano:2001qw,Cirigliano:2002jy}. As we will see in the following section, our result also agrees with the large-$N_C$ estimate, and with the value obtained from a fit to the lattice data.

\section{\boldmath Interplay with \texorpdfstring{$K\rightarrow \pi\pi$}{Kpipi} transitions}
\label{sec:latt}

As we have seen in Section \ref{sec:DStransitions}, the $\Delta S=1$ four-quark operators in Eq.~(\ref{eq:DS-4quark-Operators})
induce contributions to the corresponding LO  $\chi$PT Lagrangian in Eq.~(\ref{eq:LO_DS_ChPT_L}), which are regulated by the couplings $a_i(\mu)$. This fully determines the $K\rightarrow\pi\pi$ matrix elements at $\mathcal{O}(p^2)$. Adopting the conventions of Ref.~\cite{Cirigliano:2003gt}, the associated $\Delta I=\frac{1}{2}$ and $\Delta I=\frac{3}{2}$ decay amplitudes induced by the operator $Q_i$ are easily found to be:
\begin{align}\label{eq:LO-Kpp}
\mathcal{A}_{1/2}^{Q_{i}}(\mu)\, &\equiv\, \langle Q_i\rangle_{1/2}\, =\,
\left(\frac{1}{9}\, g_{27}^{Q_{i}}(\mu)+g_{8}^{Q_{i}}(\mu)\right) \sqrt{2}F\, (M_{K}^{2}-M_{\pi}^{2})-\frac{2\sqrt{2}}{3}\, F^{3}\, (e^2g_{8}g_{\mathrm{ewk}})^{Q_{i}}(\mu) \, ,
\nonumber\\
\mathcal{A}_{3/2}^{Q_{i}}(\mu)\, &\equiv\, \langle Q_i\rangle_{3/2}\, =\,
\frac{10}{9}\, g_{27}^{Q_{i}}(\mu)\, F\, (M_{K}^{2}-M_{\pi}^{2})
-\frac{2}{3}\, F^{3}\, (e^2g_{8}g_{\mathrm{ewk}})^{Q_{i}}(\mu)\, .
\end{align}
The factors $g_{j}^{Q_{i}}(\mu)$, which contain the $a_i(\mu)$ couplings, can be directly obtained from Eqs.~(\ref{eq:g27}) (\ref{eq:g8}) and (\ref{eq:g8gew}) by simply taking $C_{i}(\mu)=1$ and $C_{k\neq i}(\mu)=0$. 

At NLO in the chiral expansion one must take into account: 1) the different ways the LO realization of the operators $Q_i$ can be combined with the rest of the $\chi$PT building blocks to induce such a transition, and 2) new NLO building blocks with the appropriate transformation properties, which can be obtained in a similar way as it was done for the LO ones in Section~\ref{sec:4quark_EFT}. They generate the $\mathcal{O}(p^4)$ $\Delta S=1$ $\chi$PT Lagrangians of Refs. \cite{Kambor:1989tz,Ecker:1992de,Ecker:2000zr} and explicit values for their corresponding LECs $N^{Q_{i}}$, $D^{Q_{i}}$, and $Z^{Q_{i}}$ can be obtained in terms of mass-independent NLO dynamical parameters. By doing that, one can keep track of both the short-distance renormalization scale $\mu$ and the chiral scale $\nu_{\chi}$.

In the isospin limit, the NLO $K\rightarrow\pi\pi$ amplitudes induced by the set of operators $Q_i$ can be expressed in the form:
\begin{equation}\label{eq:qdelgen}
\langle Q_i\rangle_{\Delta I}\, =\,F_{\pi}\, \left\{(M_{K}^{2}-M_{\pi}^{2}) \left[A^{Q_i\, (8)}_{\Delta I}+A^{Q_i\, (27)}_{\Delta I} \right]
-F^{2}\, A^{Q_i\, (g)}_{\Delta I}\right\} \, ,
\end{equation}
with components ($X=8,27,g$)
\begin{equation}\label{eq:ampkpipi}
A^{Q_{i}\, (X)}_{\Delta I}\, =\, \mathtt{a}^{(X)}_{\Delta I}\;
g^{Q_{i}}_{X} \left[ 1+\Delta^{L}_{R} A^{(X)}_{\Delta I}+i\, \Delta^{L}_{I} A_{\Delta I}
+\Delta^C A_{\Delta I}^{Q_{i}\, (X)}
\right]
\end{equation}
where $\mathtt{a}^{(X)}_{\Delta I}$ are the tree-level normalizations in Eq.~(\ref{eq:LO-Kpp}) and $g^{Q_i}_X$ the tree-level contributions induced by $Q_i$
to the couplings  ${g}_{8}^{Q_i}$, ${g}_{27}^{Q_i}$, and $(e^2g_{8}^{\phantom{a}}g_{\mathrm{ewk}})^{Q_i}$. The dispersive and absorptive parts of the chiral loop corrections (the absorptive part fully comes from $\pi\pi$ re-scattering) are parametrized by $\Delta^{L}_{R} A^{(X)}_{\Delta I}$ and $\Delta^{L}_{I} A_{\Delta I}$, respectively, while the local counterterm contributions are included in 
$\Delta^{C}A_{\Delta I}^{Q_i\, (X)}$.
All these NLO $\chi$PT corrections can be taken from Ref.~\cite{Cirigliano:2003gt}. 

The re-scattering of the final pions generates large phase shifts in the $K\to (\pi\pi)_I$ decay amplitudes into the two possible final states with isospin $I=0$ and $2$:
\begin{equation}
  \mathcal{A}_{1/2}\, =\, A_{0}\; e^{i\chi_{0}} \, ,
  \qquad\qquad\qquad
 \mathcal{A}_{3/2}\, =\, A_{2}\; e^{i\chi_{2}} \, , 
\end{equation}
where $A_{0,2}$ are real and positive if CP is conserved. In the isospin limit, the phases $\chi_{0,2}$ can be identified with the S-wave $\pi\pi$ scattering phase shifts $\delta_{I}^0(s)$ at $s=M_{K}^{2}$ (Watson's theorem). The absorptive contributions in Eq.~(\ref{eq:ampkpipi}) are given by the tree-level amplitudes times universal corrections $\Delta^{L}_{I} A_{1/2}$ and  $\Delta^{L}_{I} A_{3/2}$, which only depend on the isospin quantum number and reproduce the $\chi$PT values of the $I=0$ and $I=2$ $\pi\pi$ phase shifts at LO in the momentum expansion, {\it i.e.}, at $\mathcal{O}(p^2)$~\cite{Pallante:1999qf}. Thus, the one-loop $\chi$PT calculation only gives the first term in the Taylor expansion of $\sin{(\delta_{I}^0)} = \delta_{I}^0 + \mathcal{O}[(\delta_{I}^0)^3]$. This implies that $\cos{(\delta_{I}^0)} = 1$ at this $\chi$PT order and, therefore, the NLO dispersive amplitudes and the moduli $A_I$ are equal up to higher-order contributions:
$A_I = \mathrm{Dis}\, (\mathcal{A}_{\Delta I}) + \mathcal{O}(p^6)$. In the limit of isospin conservation, these quantities satisfy the relation
\begin{equation}
    A_I = \mathrm{Dis}\, (\mathcal{A}_{\Delta I})\; \Theta_{\delta_I}\, ,
    \qquad\qquad
     \Theta_{\delta_I}\equiv\sqrt{1 +\tan^2{(\delta_{I}^0)}}\, .
\end{equation}
Using the LO $\chi$PT prediction for the phase shifts, this brings back the absorptive one-loop contributions that result in
$\Theta_{\delta_0} = 1.10$ and $\Theta_{\delta_2} = 1.02$. Using instead the physical values of $\delta_{I}^0(M_K^2)$ \cite{Colangelo:2001df}, which include higher-order $\chi$PT corrections, one gets 
\begin{equation}\label{eq:FSItheta}
\Theta_{\delta_0} = 1.29\pm 0.03 \, ,
\qquad\qquad\qquad
\Theta_{\delta_2} = 1.011\pm 0.004\, .
\end{equation}
This final-state-interaction effect induces a strong 30\% enhancement of the isoscalar amplitude, while the isotensor one is only modified by a mild 1\% correction~\cite{Pallante:1999qf}.

We can then extract the dispersive contributions $\mathrm{Dis}\, (\langle Q_i\rangle_{1/2})$ and $\mathrm{Dis}\, (\langle  Q_i\rangle_{3/2})$ from Eq.~(\ref{eq:qdelgen}) and obtain the corresponding isospin amplitudes $\langle Q_i\rangle_0 \equiv \mathrm{Dis}\, (\langle  Q_i\rangle_{1/2})\,\Theta_{\delta_0}$ and $\langle  Q_i\rangle_2 \equiv \mathrm{Dis}\, (\langle  Q_i\rangle_{3/2})\,\Theta_{\delta_2}$ with the correction factors in Eq.~(\ref{eq:FSItheta}), achieving a resummation of the large absorptive contributions.
All needed ingredients can be
taken from the tables of Refs. \cite{Cirigliano:2003gt,Gisbert:2017vvj,Cirigliano:2019cpi}.\footnote{Notice, however, our slightly different definition of the amplitudes $A^{Q_i\, (g)}_{\Delta I}$ that differs by a factor $F_\pi^2/F^2$ from the one adopted in Refs.~\cite{Cirigliano:2003gt,Gisbert:2017vvj,Cirigliano:2019cpi}.} At NLO in the chiral counting, 
the isoscalar amplitudes take the form:
\begin{align}\label{eq:firstq0}
\langle Q_{1}(\mu) \rangle_{0}\, &=\,\sqrt{2} F_{\pi} (M_{K}^2 -M_{\pi}^2)\,\Theta_{\delta_{0}} 
\left\{ \frac{1}{10}\left[ a_8^S(\mu) -5\,a_8^A(\mu) \right] \left(1 + \Delta^L_R A_{1/2}^{(8)} + \Delta^C_{Q_1}\right) 
\right.\nonumber \\ & \hskip 4.5cm\left.
+\, 
\frac{1}{15}\, a_{27}(\mu) \left( 1 + \Delta^L_R A_{1/2}^{(27)} +  \Delta^C_{Q_1}\right)
\right\} ,
\nonumber\\
\langle Q_{2}(\mu) \rangle_{0}\, &=\,\sqrt{2} F_{\pi} (M_{K}^2 -M_{\pi}^2)\,\Theta_{\delta_{0}} 
\left\{ \frac{1}{10}\left[ a_8^S(\mu) + 5\,a_8^A(\mu) \right] \left(1 + \Delta^L_R A_{1/2}^{(8)} +  \Delta^C_{Q_1}\right) 
\right.\nonumber \\ & \hskip 4.5cm\left.
+\, 
\frac{1}{15}\, a_{27}(\mu) \left( 1 + \Delta^L_R A_{1/2}^{(27)} +  \Delta^C_{Q_1}\right)
\right\} ,
\nonumber\\
\langle Q_{3}(\mu) \rangle_{0}\, &=\,\frac{1}{\sqrt{2}} \,  F_{\pi} (M_{K}^2 -M_{\pi}^2)\,\Theta_{\delta_{0}}
\left[ a_8^S(\mu) - a_8^A(\mu) \right]   \left( 1 + \Delta^L_R A_{1/2}^{(8)} \right) ,
\nonumber\\
\langle Q_{5}(\mu)\rangle_{0}\, &=\, 4 \sqrt{2}  F_{\pi} (M_{K}^2 - M_{\pi}^2)\,\Theta_{\delta_{0}}\, a_{LR}^{\delta\delta}(\mu) \left( 1 + \Delta^L_R A_{1/2}^{(8)} \right) ,
\nonumber\\
\langle Q_{6}(\mu) \rangle_{0}\, &=\, 4\sqrt{2} F_{\pi} (M_{K}^2 - M_{\pi}^2)\,\Theta_{\delta_{0}} \left[ 2\, a_{LR}^{\lambda\lambda}(\mu) +  \frac{1}{N_C}\, a_{LR}^{\delta\delta}(\mu) \right] 
\left( 1 + \Delta^L_R A_{1/2}^{(8)} + \Delta^C_{Q_6}\right) ,
\nonumber\\
\langle Q_{7}(\mu) \rangle_{0}\, &=\, -4 \sqrt{2}  F_{\pi}\,\Theta_{\delta_{0}}\, F^2 a_{88}^{\delta\delta}(\mu)
\left( 1 +  \Delta^L_R A_{1/2}^{(g)} \right) ,
\nonumber\\
\langle Q_{8}(\mu) \rangle_{0}\, &=\, -4 \sqrt{2} F_{\pi}\,\Theta_{\delta_{0}} \left[ 2\, F^2 a_{88}^{\lambda\lambda}(\mu) + \frac{1}{N_{c}}\, F^2 a_{88}^{\delta\delta}(\mu)\right] 
\left( 1 +  \Delta^L_R A_{1/2}^{(g)} + \Delta^C_{Q_8,0} \right)  ,
\end{align}
while the $I=2$ amplitudes are given by:
\begin{align}\label{eq:firstq2}
\langle Q_{1}(\mu) \rangle_{2}\, &=\, \frac{2}{3}\,  F_{\pi} (M_{K}^2 - M_{\pi}^2)\,\Theta_{\delta_{2}}\; a_{27}(\mu)
\left(1 + \Delta^L_R A_{3/2}^{(27)} +  \Delta^C_{Q_1} \right) ,
\nonumber\\
\langle Q_{7}(\mu) \rangle_{2}\, &=\, -4  F_{\pi}\,\Theta_{\delta_{2}}\; F^2 a_{88}^{\delta\delta}(\mu)
\left(1 + \Delta^L_R A_{3/2}^{(g)} \right) , 
\nonumber\\
\langle Q_{8}(\mu) \rangle_{2}\, &=\, -4 F_{\pi}\,\Theta_{\delta_{2}}  
\left[ 2\, F^2 a_{88}^{\lambda\lambda}(\mu) + \frac{1}{N_c}\, F^2a_{88}^{\delta\delta}(\mu)\right] 
\left(1 + \Delta^L_R A_{3/2}^{(g)} + \Delta^C_{Q_8,2} \right)  .
\end{align}
The local counterterm contributions have been approximated by their large-$N_{C}$ expressions~\cite{Cirigliano:2003gt,Gisbert:2017vvj,Cirigliano:2019cpi}:
\begin{align}\label{eq:DeltasC}
\Delta^C_{Q_1}\, &=\, \frac{4 M_\pi^2}{F_\pi^2}\, L_5\, ,
\nonumber\\
\Delta^C_{Q_6}\, &=\, \frac{4 M_K^2}{F_\pi^2}\, \left[ 2 L_8 -\frac{1}{4}\,  L_5\, ( 1-16 \lambda_3^{SS})\right]
+ \frac{4 M_\pi^2}{F_\pi^2}\, \left[ \frac{8 L_8^2}{L_5} - L_5\, (3 + 4 \lambda_3^{SS})
\right] 
+\mathcal{O}(\bar\lambda^{RR}_i)\, ,
\nonumber\\
\Delta^C_{Q_8,0}\, &=\, \frac{4 M_K^2}{F_\pi^2}\, (4 L_8 - L_5) + \frac{16 M_\pi^2}{F_\pi^2}\, (2 L_8 -  L_5)\, ,
\nonumber\\
\Delta^C_{Q_8,2}\, &=\, \frac{8 M_K^2}{F_\pi^2}\, (2 L_8 - L_5) + \frac{4 M_\pi^2}{F_\pi^2}\, (8 L_8 - 3 L_5)\, .
\end{align}
The numerical values of the different loop and counterterm corrections are given in Tables~\ref{tab:numerics1} and~\ref{tab:numerics2}. The uncertainties quoted for the loop contributions have been estimated by varying the chiral scale $\nu_{\chi}$ in the interval $(0.6-1.0)$ GeV. To estimate the smaller counterterm contributions, we have used the same input values for the $\chi$PT LECs than Ref.~\cite{Cirigliano:2019cpi}; their associated parametric uncertainties are reflected in the errors displayed in  Table~\ref{tab:numerics2}.

\begin{table}[tbh]
\centering
\begin{tabular}{|c|c|c|c|c|c|}
\hline
$\Delta^L_R A_{1/2}^{(8)}$ & $\Delta^L_R A_{1/2}^{(27)}$ & $\Delta^L_R A_{1/2}^{(g)}$
& $\Delta^L_R A_{3/2}^{(27)}$ & $\Delta^L_R A_{3/2}^{(g)}$
\\ \hline
 $0.27 \pm 0.05$ & $1.02 \pm 0.63$ & $0.44 \pm 0.10$ & $0.01 \pm 0.05$ & $-0.34\pm 0.10$ \\ \hline
\end{tabular}
\caption{Numerical values for the dispersive loop amplitudes $\Delta^L_R A_{\Delta I}^{(X)}$. 
}
\label{tab:numerics1}
\end{table}

\begin{table}[tbh]
\centering
\begin{tabular}{|c|c|c|c|}
\hline
$\Delta^C_{Q_1}$ & $\Delta^C_{Q_6}$       & $\Delta^C_{Q_8,0}$      & $\Delta^C_{Q_8,2}$             \\ \hline
$0.010 \pm 0.001$ & 
$0.15 \pm 0.03$ & 
$0.10 \pm 0.06$ & 
$-0.03 \pm 0.06$  \\ \hline
\end{tabular}
\caption{Numerical values for the counterterm contributions.} 
\label{tab:numerics2}
\end{table}

The matrix elements of $Q_4$, $Q_9$ and $Q_{10}$ are not independent because of the relations among operators given in footnote~\ref{foot:OpRel}. Thus,
\begin{align}
\langle Q_{4} \rangle_{0}\, &=\, -\langle Q_{1} \rangle_{0}+\langle Q_{2} \rangle_{0}+\langle Q_{3} \rangle_{0} \, ,
\nonumber\\
\langle Q_{9} \rangle_{0}\, &=\, \frac{3}{2}\,\langle Q_{1} \rangle_{0}-\frac{1}{2}\,\langle Q_{3} \rangle_{0} \, , 
\nonumber\\
\langle Q_{10} \rangle_{0}\, &=\, \frac{1}{2}\,\langle Q_{1} \rangle_{0}+ \langle Q_{2} \rangle_{0}-\frac{1}{2}\,\langle Q_{3} \rangle_{0} \, .
\label{eq:lastq}
\end{align}
Notice that the strong penguin operators $Q_{3,4,5,6}$ cannot induce a $\Delta I=\frac{3}{2}$ transition and, therefore, their corresponding matrix elements into an $I=2$ $\pi\pi$ final state are identically zero. Moreover, isospin symmetry implies
\begin{equation}
 \langle Q_{1} \rangle_{2}\, =\, \langle Q_{2} \rangle_{2}\, =\,\frac{2}{3}\,\langle Q_{9} \rangle_{2}\, =\,\frac{2}{3}\,\langle Q_{10} \rangle_{2} \, .   
 \label{eq:I2relations}
\end{equation}

Using the theoretically-estimated value of $a_{27}(\mu_0)$ in Eq.~(\ref{eq:a27det}), one finds 
\begin{equation}
\langle Q_{1}(\mu_0) \rangle_{2}\, =\, 0.0058\; (23)_{a_{27}}(3)_{\Delta_{L}} \, ,
\end{equation}
in the $\overline{\mathrm{MS}}$-NDR scheme at $\mu_0=1\, \mathrm{GeV}$, where the first uncertainty is the parametric error from $a_{27}(\mu_0)$ and the second one accounts for missed subleading chiral corrections.
The CP-conserving part of the amplitude $A_2$ is totally dominated by the contributions from the operators  $Q_{1}$ and $Q_{2}$. Taking the corresponding Wilson Coefficients from Table~\ref{tab:wil}, one then predicts:
\begin{equation}
\mathrm{Re}\, (A_{2})^{\mathrm{th}}\, =\, \left( 0.82\pm 0.33\right) \cdot 10^{-8}\;\mathrm{GeV} \, ,
\end{equation}
in reasonable agreement with the experimental value $\mathrm{Re}\, (A_{2})^{\mathrm{exp}}=1.210\, (2)\cdot 10^{-8}$~GeV \cite{Cirigliano:2011ny}. The measured value of $A_2$ is of course exactly reproduced, taking instead as input the phenomenological determination of $a_{27}(\mu_0)$ in Eq.~(\ref{eq:a27-exp}).

From the measured $\tau$ spectral functions, we have been able to determine $F^2 a_{88}^{\lambda\lambda}(m_\tau)$ in Eq.~(\ref{eq:a88det}), which allows us to predict the $K\to \pi\pi$ matrix elements of the operator $Q_8$. Safely neglecting the very suppressed $a_{88}^{\delta\delta}$ contribution, we find at $\mu_0=1\, \mathrm{GeV}$:
\begin{align}
\langle Q_{8}(\mu_0) \rangle_{0}\, &=\, 1.62\; (45)_{a_{88}^{\lambda\lambda}}(10)_{\Delta_{L}}(6)_{\Delta_{C}} \, ,
\nonumber\\
\langle Q_{8}(\mu_0) \rangle_{2}\, &=\, 0.37\; (10)_{a_{88}^{\lambda\lambda}}(6)_{\Delta_{L}}(3)_{\Delta_{C}} \, .
\label{eq:Q82val}
\end{align}
The isotensor matrix element governs the SM contribution to the CP-violating ratio $\varepsilon'/\varepsilon$  associated with the ($I=2$) electroweak penguin operators \cite{Gisbert:2017vvj}, 
\begin{equation}
(\varepsilon'/\varepsilon)_{\mathrm{EWP}}^{(2)}\,\equiv\,\frac{1}{\sqrt{2}|\varepsilon|}\,\frac{\mathrm{Im}\, (A_{2})^{\mathrm{EWP}}}{\mathrm{Re}\, (A_{0})} \, ,
\end{equation}
which at $\mu=1\, \mathrm{GeV}$ is dominated by $Q_{8}$. 
Taking the experimental values of  $\mathrm{Re}\, (A_{0})^{\mathrm{exp}}= 2.704\, (1)\cdot 10^{-7}$~GeV \cite{Cirigliano:2011ny}
and $|\varepsilon|^{\mathrm{exp}} = 2.228\, (11)\cdot 10^{-3}$ \cite{Zyla:2020zbs}, and
\begin{equation}
\mathrm{Im}\, (A_{2})^{\mathrm{EWP}}_{Q_{8}}\, =\, G\; \mathrm{Im}\, [C_{8}(\mu)]\;\langle Q_{8}(\mu) \rangle_{2} \;  ,
\end{equation}
one finds
\begin{equation}
(\varepsilon'/\varepsilon)^{(2)}_{\mathrm{EWP},Q_{8}}\, =\, \left(-5.6\pm 1.5_{\, a_{88}^{\lambda\lambda}}\pm 0.9_{\Delta_{L}}\pm 0.5_{\Delta_{C}}\right)\cdot 10^{-4}=(-5.6\pm 1.8) \cdot 10^{-4}\, .
\end{equation}
On the other hand, using Eq.~(\ref{eq:I2relations}) the (smaller) $Q_{9,10}$ contribution is simply given by
\begin{equation}
(\varepsilon'/\varepsilon)^{(2)}_{\mathrm{EWP},Q_{9,10}}\, =\,\frac{3\omega}{2\sqrt{2}|\varepsilon|}\,\frac{\mathrm{Im}(C_{9}+C_{10})}{\mathrm{Re}(C_{1}+C_{2})}\, =\, (1.1\pm 0.1)\cdot 10^{-4} \, ,
\end{equation}
where the ratio $\omega\equiv \mathrm{Re}A_{2}/\mathrm{Re}A_{0} = 0.0447\, (1)$ has been taken from experimental data. Adding this contribution one finally finds

\begin{equation}
(\varepsilon'/\varepsilon)^{(2)}_{\mathrm{EWP}}\, =\, \left(-4.5\pm 1.5_{\, a_{88}^{\lambda\lambda}}\pm 0.9_{\Delta_{L}}\pm 0.5_{\Delta_{C}}\right)\cdot 10^{-4}=(-4.5\pm 1.8) \cdot 10^{-4}\, .
\end{equation}
This result agrees very well with the value $-(3.5 \pm 2.2)\cdot  10^{-4}$, obtained in Refs.~\cite{Gisbert:2017vvj,Cirigliano:2019cpi} with a large-$N_C$ estimate of $a_{88}^{\lambda\lambda}$ (as well as the smaller contributions of the other couplings),\footnote{We thank Hector Gisbert for cross-checking this number.}
instead of our determination from $\tau$ decay data.

\subsection{Fit to lattice data}
\label{subsec:lattice}

Our NLO results for the kaon decay amplitudes allow us to perform a direct fit to the lattice data of the RBC-UKQCD collaboration~\cite{Abbott:2020hxn}. The  numerical values for the matrix elements of the different four-quark operators provided in Ref.~\cite{Abbott:2020hxn} can be fitted to our analytic expressions in Eqs.~(\ref{eq:firstq0}),~(\ref{eq:firstq2}),~(\ref{eq:lastq}) and~(\ref{eq:I2relations}).
In Ref.~\cite{Abbott:2020hxn}, the ten $I=0$ matrix elements are given at $\mu=4\, \mathrm{GeV}$ in the $\overline{\mathrm{MS}}$ scheme, together with their statistical covariance matrix.\footnote{A useful comparison of the different normalization conventions can be found in Ref.~\cite{Aebischer:2020jto}.} Systematic uncertainties are estimated to be a $15.7 \%$. We run those matrix elements to $\mu=1\, \mathrm{GeV}$, propagating their uncertainties, and use afterwards the relations (\ref{eq:lastq}) to reduce the operator basis to the seven independent $I=0$ operators. The matrix elements of the three independent (in the isospin limit) $I=2$ operators can also be found in Ref.~\cite{Abbott:2020hxn} (see also Refs. \cite{Blum:2015ywa} and \cite{Blum:2012uk}).

The fitted results for our seven $a_i(\mu)$ parameters are displayed in Table \ref{tab:fit}. The fit returns a relatively small $p$-value ($p= 8 \%$), which mainly arises from a small tension between  $\langle Q_{8} \rangle_0$ and $\langle Q_{8} \rangle_2$ (the lattice determination of $\langle Q_{8} \rangle_0$ favours smaller values for $|a_{88}^{\lambda\lambda}|$ than $\langle Q_{8} \rangle_2$). The fitted parameters are in good agreement with the phenomenological values of $a_{27}^{\phantom{S}}$ and $F^2 a_{88}^{\lambda\lambda}$ found in the previous sections, which are shown in the second line of the table. 
The third line collects the predicted numerical values for those couplings in the large-$N_{C}$ limit, given in 
Section~\ref{subsec:LargeNc}. 
This limit is able to correctly reproduce the hierarchy of the couplings, with the exception of $a_{27}$ and, especially, $a_{8}^{A}$. 
Notice also the large error in the fitted value of the coupling $a_{8}^{S}(\mu_0)$ that governs the contribution of the operator $Q_+\equiv Q_2+Q_1$ to the isoscalar $K\to\pi\pi$ amplitude. With the current precision, the lattice data are still insensitive to this parameter because its contribution to $g_8$ in Eq.~(\ref{eq:g8}) is suppressed by a factor $1/10$.

\setlength{\tabcolsep}{3pt}
\renewcommand{\arraystretch}{1.2}
\begin{table}[tbh]
\begin{small}
\begin{tabular}{|l|l|l|l|l|l|l|l|}
\hline
& $a_{27}^{\phantom{S}}(\mu_0)$    & $a_{8}^{S}(\mu_0)$  & $a_{8}^{A}(\mu_0)$ & $a_{LR}^{\delta\delta}(\mu_0)$ & $a_{LR}^{\lambda\lambda}(\mu_0)$ & $F^2 a_{88}^{\delta\delta}(\mu_0)$ & $F^2 a_{88}^{\lambda\lambda}(\mu_0)$ \\ \hline
Lattice   & $0.64\, (10)$ & $-0.2\, (24)$ & $2.7\, (5)$ & $-0.48\, (41)$            & $-1.17\, (26)$              & $-0.22\, (12)$                      & $-0.68\, (11)\;\mathrm{GeV}^2$                            \\ \hline
$K,\tau$ data 
& $0.622\, (43)$   &      &    &                    &                  &                           & $-0.78\, (22)\;\mathrm{GeV}^2$                                \\ \hline
Large $N_{C}$ & $1$     & $1$     & $1$   & $0$                    & $-1.06$                 & $0$                            & $-0.70\;\mathrm{GeV}^2$                                \\ \hline
\end{tabular}
\end{small}
\caption{\label{tab:fit} Values at $\mu_0=1\, \mathrm{GeV}$ ($\overline{\mathrm{MS}}$-NDR) of the $a_i(\mu_0)$ parameters, extracted from a  NLO fit to the lattice data (first line) and from experimental data (second line), compared with their large-$N_{C}$ predictions (third line).}
\end{table}

From the measured $K\to\pi\pi$ rates, it is not possible to extract separate values for the different octet couplings. The experimental data only determines the combination $g_8$ in Eq.~(\ref{eq:g8}). Taking into account the absorptive resummation factor $\Theta_{\delta_0}$ in Eq.~(\ref{eq:FSItheta}),\footnote{
Including only the absorptive one-loop contribution, {\it i.e.} with $\Theta_{\delta_0}= 1.10$, one gets instead 
$g_8^{\mathrm{exp}}\, =\, 3.60\pm 0.14$~\cite{Cirigliano:2019cpi}.
}
one obtains
\begin{equation}\label{eq:g8exp}
    g_8^{\mathrm{exp}}\, =\, 3.07\pm 0.14\, .
\end{equation}
Our fit to the lattice data implies $g_8^{\mathrm{Latt}}\, =\, 2.6\pm 0.5$, in good agreement with (\ref{eq:g8exp}), while the large-$N_C$ determination of the $a_i$ couplings gives a value $g_8^{\infty}\, =\, 1.2\pm 0.4$ that is clearly too small. 

The comparison between the values of the $a_i$ parameters extracted from the lattice data and their large-$N_C$ predictions provides an enlightening anatomy of the well-known $\Delta I =\frac{1}{2}$ rule in non-leptonic kaon decays. The large difference between the isoscalar and isotensor decay amplitudes results from the combination of several interrelated dynamical effects:
\begin{enumerate}
    \item The table exhibits a large enhancement of  $a_{8}^{A}(\mu_0)$ by a factor $2.7$ that complements the short-distance gluonic enhancement of $C_-(\mu_0)\equiv (C_2-C_1)(\mu_0)$ at LO \cite{Altarelli:1974exa,Gaillard:1974nj} and NLO \cite{Buras:1991jm,Buras:1992tc,Buras:1992zv,Ciuchini:1992tj,Ciuchini:1993vr}. This clearly identifies the main origin of the isoscalar enhancement in the $K\to\pi\pi$ matrix element of the operator $Q_-\equiv Q_2-Q_1$, confirming the findings of many previous approaches \cite{Bardeen:1986vz,Pich:1990mw,Pich:1995qp,Antonelli:1995gw,Antonelli:1995nv,Bertolini:1997ir,Bijnens:1998ee,Hambye:1999ic,Buras:2014maa}. 
    
    \item The matrix element of the penguin operator $Q_6$ receives a chiral enhancement through the factor $8\, a_{LR}^{\lambda\lambda}(\mu_0)$. In spite of the small numerical value of the Wilson coefficient $C_6(\mu_0)$, this provides an additional ($\sim 10\%$ at $\mu=1$~GeV) increment of the $I=0$ amplitude \cite{Vainshtein:1975sv,Shifman:1975tn}. 
    Since the anomalous dimension of $Q_6$ is leading in $1/N_C$, the large-$N_C$ limit is able to capture the chiral enhancement factor, providing a very good approximation to $a_{LR}^{\lambda\lambda}(\mu_0)$, as exhibited in Table~\ref{tab:fit}. However, this is not enough to reproduce the physical hadronic matrix element of $Q_6$ \cite{Hambye:2003cy}. One still needs to incorporate the very sizeable corrections from $\chi$PT loops \cite{Pallante:1999qf}.
    
    \item The $\chi$PT loop contributions are subleading in the $1/N_C$ counting but they are enhanced by large infrared logarithms and, moreover, contain very important unitarity corrections associated with the final-state interactions of the emerging pions \cite{Pallante:1999qf}.
    As shown in Table~\ref{tab:numerics1}, the one-loop $\chi$PT correction provides a sizeable 30\% enhancement of the isoscalar amplitude \cite{Kambor:1991ah,Pallante:1999qf,Pallante:2000hk,Pallante:2001he} that is further reinforced by the all-order resummation of absorptive contributions through the factor $\Theta_{\delta_0}$ in Eq.~(\ref{eq:FSItheta}). The corresponding $\chi$PT corrections on 
    $\mathrm{Re}(A_2)$ are very mild.
    
    \item In addition, there is a sizeable suppression of $a_{27}^{\phantom{S}}(\mu_0)$ by about $30-40\% $, with respect to its expected value at $N_C\to\infty$, which implies a corresponding suppression of the amplitude $A_2$. This effect was suggested long time ago through a large-$N_C$ topological analysis of the $K\to\pi\pi$ amplitudes \cite{Pich:1995qp}, showing that the leading and subleading contributions in $1/N_C$ (excluding penguins) appear anticorrelated in $g_8$ and $g_{27}$, so that the enhancement of one coupling requires the suppression of the other.\footnote{
 At LO, the topological parameters $a,b,c$ defined in Ref.~\cite{Pich:1995qp} can be easily related to our $a_i(\mu)$ couplings: 
    $a+b = a_{27}\, (C_1+ C_2)$, \ 
    $b\approx \frac{1}{2}\, a_8^A\, (C_1-C_2+C_3) + \frac{1}{10}\, (6\, a_{27} - a_8^S)\, (C_1+C_2) - \frac{1}{2}\, a_8^S\, C_3$
    \ and \
    $c\approx 8\, a_{LR}^{\lambda\lambda}\, C_6 + \frac{1}{2}\, (a_8^A+a_8^S)\, C_4$.
    They are also directly related to the lattice topologies discussed in Ref.~\cite{Boyle:2012ys}.
}
    The anticorrelation  of the two colour structures has been numerically confirmed by the RBC-UKQCD lattice evaluation of $A_2$ \cite{Boyle:2012ys}, and corroborated by a more recent lattice analysis of the scaling with $N_C$ of the $K\to\pi$ amplitudes in a simplified setting with four degenerate quark flavours ($m_u=m_d=m_s=m_c$) \cite{Donini:2016lwz,Donini:2020qfu}.
    
\end{enumerate}

It is worth mentioning at this point that these dynamical features are fully supported at the inclusive level by the NLO calculation of the two-point correlation function (without electroweak penguin operators)
\begin{equation}
    \Psi(q^2)\,\equiv\,\int d^4x\; \mathrm{e}^{iqx}\;\langle 0| T(\mathcal{L}^{\Delta S=1}(x)\, \mathcal{L}^{\Delta S=1}(0)^\dagger) |0\rangle\, =\, 
    G^2 \sum_{i,j=1}^6\, C_i(\mu)\, C_j(\mu)^*\;\psi_{ij}(q^2)\, ,
\end{equation}
presented in Refs.~\cite{Pich:1988qe,Pich:1990mw,Jamin:1994sv,Pich:1995qp}. This correlator does not involve any hadronic state and, therefore, can be rigorously analyzed with short-distance QCD methods. In order to better visualise the large impact of gluonic corrections, it is convenient to simplify the discussion and restrict ourselves to the non-penguin operators $Q_\pm$. In the absence of penguin-like contributions, these two operators are multiplicatively renormalizable, which allows one to derive compact analytical expressions for the
spectral functions associated with the $C_\pm(\mu)\, Q_\pm$ terms (exact numerical results for the full correlator can be found in Ref.~\cite{Jamin:1994sv}):
\begin{align}\nonumber
\rho_\pm (t) &\equiv \frac{1}{\pi}\,\mathrm{Im}\Psi_{\pm\pm}(t)\, \\&=\, \theta(t)\, \frac{2}{45}\, G^2\, C^2_\pm(M_W^2)\, N_C^2\left(1\pm\frac{1}{N_C}\right)\frac{t^4}{(4\pi)^6}\,\alpha_s(t)^{-2\hat\gamma_\pm}\left[ 1 + \zeta_\pm\,\frac{\alpha_s(t)}{\pi}\right] ,\label{eq:Psipm}
\end{align}
where the $\frac{1}{N_C}$-suppressed powers $\hat\gamma_\pm =\gamma_\pm^{(1)}/\beta_1 = \pm \frac{9}{11 N_C}\, (1\mp \frac{1}{N_C})/(1-\frac{6}{11 N_C})$ contain the LO anomalous dimensions that enhance the Wilson coefficient $C_-(\mu)$ ($\hat\gamma_- = -\frac{4}{9}$) and suppress $C_+(\mu)$ ($\hat\gamma_+ = +\frac{2}{9}$). Since $\Psi(t)$ is a renormalization-invariant quantity, the logarithmic $\alpha_s$ corrections have been already reabsorbed with the choice $\mu^2=t$. At this level of approximation ($\zeta_\pm = 0$), 
it is impossible to understand the big ratio $A_0/A_2$  (or, equivalently, $g_8/g_{27}$) with the information provided by
the spectral functions $\rho_\pm(t)$ \cite{Pich:1985st,Pich:1987sc}. The physics picture gets completely changed once the NLO corrections are included: $\rho_-(t)$ gets a huge enhancement through the positive NLO correction $\zeta_- = \frac{9139}{810}$, while the corresponding correction to $\rho_+(t)$ is negative and 6 times smaller, $\zeta_+ = -\frac{3649}{1620}$~\cite{Jamin:1994sv,Pich:1995qp}. In both cases, the NLO short-distance Wilson coefficients only contribute a small part of the $\zeta_\pm$ corrections (17\% and 8\%, respectively, for $\zeta_-$
and $\zeta_+$). More interesting, this enhancement/suppression pattern completely disappears in the large-$N_C$ limit where $\zeta_+^\infty =\zeta_-^\infty = \frac{9}{4}$~\cite{Pich:1988qe}.

Since $Q_6$ is the only operator (excluding electroweak penguins) with a non-vanishing anomalous dimension at $N_C\to\infty$, it is possible to make an analogous computation of $\rho_6(t)\equiv \frac{1}{\pi}\,\mathrm{Im}\Psi_{66}(t)$ in the large-$N_C$ limit \cite{Pich:1988qe}. The result is in fact known to NNLO \cite{Pich:1990mw}: 
\begin{equation}\label{eq:Psipm6}
\rho_6 (t) \, =\, \theta(t)\, \frac{12}{5}\, G^2\, |C_6(M_W^2)|^2\, \frac{t^4}{(4\pi)^6}\,\alpha_s(t)^{18/11}\left[ 1 + \frac{117501}{4840}\,\frac{\alpha_s(t)}{\pi} + 470.72 \left(\frac{\alpha_s(t)}{\pi}\right)^2\right] .
\end{equation}
This exhibits again a huge dynamical enhancement which persists at higher perturbative orders, but this time the enhancement is already captured in the large-$N_C$ limit. The NLO Wilson coefficient only contributes a 13\% of the non-logarithmic $\mathcal{O}(\alpha_s)$ correction.

\subsection{\boldmath \texorpdfstring{$F_\pi$}{Fpi} determination 
from inclusive \texorpdfstring{$\tau$}{tau}-decay data}

\begin{figure}[tb]\centering
\includegraphics[width=1.\textwidth]{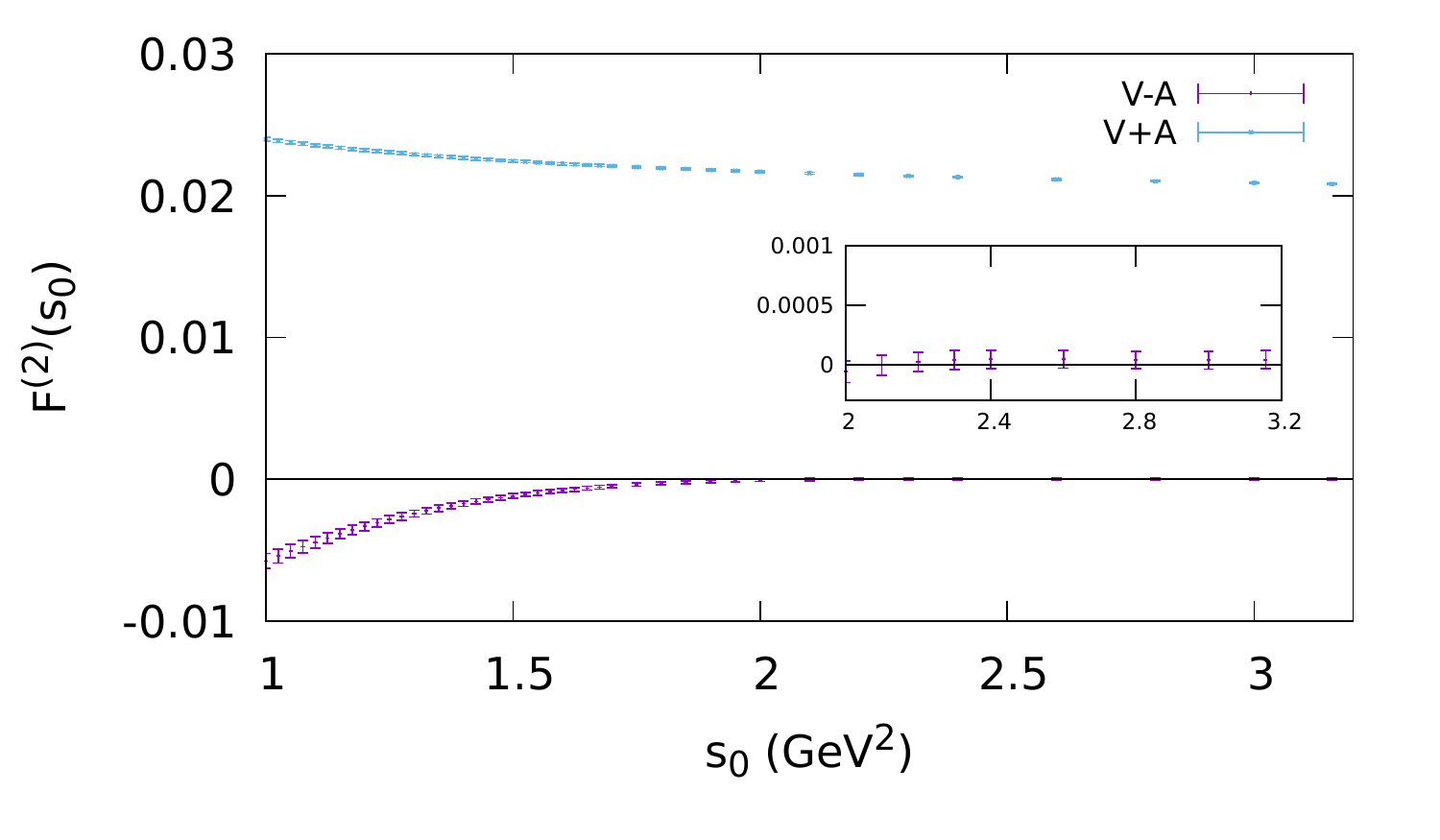}
\caption{$F^{(2)}_{V\pm A}(s_{0})$ as defined in Eq.~(\ref{eq:sumruleprecise}). The error bars include the experimental uncertainties and the theoretical errors associated with $\langle\mathcal{O}_{6,V-A}^d\rangle'$.}
\label{fig:sr2p}
\end{figure}

Instead of determining $a_{88}^{\lambda\lambda}$ from $\tau$ decays, we can use the value extracted from our fit to the lattice data
of the RBC-UKQCD collaboration.
Since we have also fitted $a_{88}^{\delta\delta}$, we can obtain the full dimension-six contribution to the OPE of $\Pi^d_{V-A}(s)$, at NLO in both $\alpha_{s}$ and the $\chi$PT expansion. 
Taking into account the complete scale dependence of $\langle\mathcal{O}^d_{6,V-A}(\mu)\rangle$ in Eq.~(\ref{eq:fullcond}), the dispersion relation of Eq.~(\ref{sumrule22}) for the weight $\hat\omega(s)=\left(1- s/s_{0}\right)^2$ generalizes, up to corrections suppressed  by $\alpha_{s}$ and $8$ powers of the energy scale, to
\begin{align}
\int^{s_{0}}_{s_{\mathrm{th}}}\frac{ds}{s_{0}} \left(1-\frac{s}{s_{0}}\right)^2\frac{1}{\pi}\operatorname{Im}\Pi_{V-A}(s)-
\frac{\langle\mathcal{O}^{d}_{6,V-A}(s_{0})\rangle'}{s_{0}^{3}} 
+ \delta_{\mathrm{DV}}[\hat\omega(s),s_0]
\; =\; 2\, \frac{F_{\pi}^{2}}{s_{0}}\,\left(1-\frac{M_{\pi}^{2}}{s_{0}}\right)^2  ,
\end{align}
where $\langle\mathcal{O}_{6,V-A}^d(s_{0})\rangle'$ equals $\langle\mathcal{O}_{6,V-A}^d(s_{0})\rangle$ in  Eq.~(\ref{eq:fullcond}), with the changes $A_{1,8}\to A_{1,8}+ \frac{3}{2}\, B_{1,8}$ and $B_{1,8}\to 0$.
In Table~\ref{tab:fit}, the parameters $a_{88}^{\lambda\lambda}(\mu)$ and $a_{88}^{\delta\delta}(\mu)$ have been determined at $\mu=1\, \mathrm{GeV}$.  Their running up to $s_{0}$ is governed by the known $\mu$ dependence of $Q_{7}$ and $Q_{8}$ at NLO because 
the $\chi$PT coupling $e^2g_8 g_{\mathrm{ewk}}$ in Eq.~(\ref{eq:g8gew}) does not depend on the short-distance renormalization scale.
At $s_{0}=m_{\tau}^2$ one finds:

\begin{equation}
\langle\mathcal{O}_{6,V-A}^d(m_{\tau}^{2})\rangle'\, =\, -(2.9 \pm 0.5) \cdot 10^{-3} \, \mathrm{GeV} \, .
\label{eq:O6_V-A_Lattice}
\end{equation} 
The negligible role of duality violations for this weight function at $s_{0}\sim m_{\tau}^{2}$, together with the good knowledge of the very small power corrections involved, translate into a very powerful prediction for its associated integral. In Fig.~\ref{fig:sr2p} we display the $s_0$ dependence of
\begin{align}
F^{(2)}_{V\pm A}(s_{0})&\equiv \int^{s_{0}}_{s_{\mathrm{th}}}\frac{ds}{s_{0}}  \left(1-\frac{s}{s_{0}} \right)^2\frac{1}{\pi}\operatorname{Im}\Pi_{V\pm A}(s)
\pm 2\,\frac{F_{\pi}^{2}}{s_{0}}\left(1-\frac{M_{\pi}^{2}}{s_{0}}\right)^2 -\frac{\langle\mathcal{O}_{6,V\pm A}^d(s_{0})\rangle'}{s_{0}^{3}} \, .
\label{eq:sumruleprecise}
\end{align}
Similarly to what we did before in Fig.~\ref{fig:wsrmixed} with the weight $(1-s/s_0)$,
we plot also the corresponding $V+A$ integral, although neglecting in that case 
the relatively very small contribution from $\langle\mathcal{O}_{6,V+A}^d\rangle'$ that is irrelevant for the comparison.
For the $V-A$ distribution, we have used the value of $\langle\mathcal{O}_{6,V-A}^d(m_{\tau}^{2})\rangle'$ in Eq.~(\ref{eq:O6_V-A_Lattice}), running it down to every $s_{0}$ at NLO in QCD.
Above $2\;\mathrm{GeV}^2$, one observes an exact cancellation of the vector and axial-vector contributions to $F^{(2)}_{V- A}(s_{0})$,  which remains compatible with zero within $1 \sigma$, even when the experimental data are precise enough to resolve the predicted zero of $F_{V-A}^{(2)}(s_{0})$ with a $\sim 0.5\%$ accuracy with respect to the normalization of the total $V+A$ distribution. 

Since the strong cancellation involves the pion decay constant, one can exploit the theoretical prediction $F^{(2)}_{V-A}(s_{0}\sim m_{\tau}^{2})=0$ to determine $F_{\pi}$. Although the pion contribution in Eq.~(\ref{eq:sumruleprecise})  is suppressed by two powers of energy, the sensitivity is good enough to derive a precise value for $F_\pi$:
\begin{equation}\label{eq:fpiTau}
F_{\pi}^{\mathrm{incl}}=(92.6 \pm 0.6) \; \mathrm{MeV} \, ,
\qquad\qquad\qquad
\sqrt{2}\, F_{\pi}^{\mathrm{incl}}=(130.9 \pm 0.8) \; \mathrm{MeV} \, ,
\end{equation}
in perfect agreement with the values found in the literature from other sectors \cite{Zyla:2020zbs,Aoki:2019cca}. Notice that we have not used any information from the decay $\tau^{-}\rightarrow \pi^{-}\nu_{\tau}$.

Another possible application of this result is reinterpreting it as a powerful constraint on hypothetical new physics contributions that do not respect chiral symmetry at short distances. Contributions of this type would easily spoil the strong cancellation between the vector and axial-vector integrated distributions, in disagreement with the behaviour displayed in Fig.~\ref{fig:sr2p}. This idea was already exploited in Ref.~\cite{Cirigliano:2018dyk}, where powerful bounds on new physics above the $\mathrm{TeV}$ scale were extracted. 

Finally, 
we can also estimate the dimension-8 condensate,
using the triple-pinched dispersion relation
\begin{equation}
\int^{s_{0}}_{s_{th}}\frac{ds}{s_{0}}\left(1-\frac{s}{s_{0}} \right)^{3}\frac{1}{\pi}\mathrm{Im}\Pi_{V-A}(s)
-2\,\frac{F_{\pi}^{2}}{s_{0}}\left(1-\frac{m_{\pi}^{2}}{s_{0}}\right)^3
- 3\,\frac{\langle\mathcal{O}_{6,V- A}^d(s_{0})\rangle''}{s_{0}^{3}}
-\frac{\langle\mathcal{O}_{8,V-A}^d\rangle}{s_{0}^4}\, =\, 0\, ,
\end{equation}
where now $\langle\mathcal{O}_{6,V-A}^d(s_{0})\rangle''$ equals $\langle\mathcal{O}_{6,V-A}^d(s_{0})\rangle$ in  Eq.~(\ref{eq:fullcond}), with the changes 
$A_{1,8}\to A_{1,8}+ \frac{1}{2}\, B_{1,8}$    
and $B_{1,8}\to 0$, and the  equality holds up to very tiny logarithmic ($\alpha_{s}$-suppressed) corrections to $D\geq 8$ and duality violations. Since $\langle\mathcal{O}_{6,V-A}^d(s_{0})\rangle''$ is determined by our lattice fit, the $\tau$ data now implies:
\begin{equation}\label{eq:Q8VmA}
\langle\mathcal{O}_{8,V-A}^d\rangle\, =\, (-1.3 \pm 0.7)\cdot 10^{-2}\, \mathrm{GeV}^{8} \, ,
\end{equation}
which is in good agreement with the different determinations found in the literature~\cite{Rodriguez-Sanchez:2016jvw,Boito:2015fra}.

\section{Conclusions}
\label{sec:conclusions}

We have presented a detailed analysis of light-quark four-fermion operators, using the symmetry relations emerging from their chiral $SU(3)_L\otimes SU(3)_R$ structure and a low-energy effective Lagrangian approach. This has allowed us to derive rigorous relations between non-perturbative parameters appearing in different physical processes. In particular, we have studied in a systematic way the relations between
the dimension-six vacuum condensates entering the OPE of the vector and axial-vector QCD currents, and the hadronic matrix elements of weak operators in $\Delta S=1$ ($K\to\pi\pi$) and $\Delta S=2$ ($K^0-\bar K^0$) transitions. The $\chi$PT framework provides a powerful way to determine the low-energy realization of the four-quark operators, taking into account their different decomposition in irreducible representations of the chiral group and ordering their phenomenological impact through the chiral momentum expansion. The non-trivial dynamical information gets encoded in a few low-energy constants that characterize the different structures allowed by symmetry. These constants can be easily estimated in the limit of a large number of QCD colours, which provides useful reference values to compare with.

As a first important phenomenological application, we have determined the electromagnetic penguin contribution to the ratio $\varepsilon'/\varepsilon$, which parametrizes the direct violation of CP symmetry in the $K\to\pi\pi$ amplitudes. The relevant operator has an $(88)$ structure that gives rise to a leading $\mathcal{O}(p^0)$ contribution, providing a sizeable chiral enhancement of its matrix elements. The symmetry relations connect this $\mathcal{O}(p^0)$ term with the vacuum matrix element of the corresponding four-quark operator
appearing in the OPE of the $\Pi^d_{V-A}$ correlator, which is accessible through hadronic $\tau$ decay data. Using the measured invariant-mass distribution of the final hadrons in $\tau$ decays, we have found 
\begin{equation}
(\varepsilon'/\varepsilon)^{(2)}_{\mathrm{EWP}}=-(4.5\pm 1.8)\cdot 10^{-4} \, ,
\end{equation}
at NLO in $\chi$PT. This phenomenological determination is in excellent agreement with the values obtained in the $\chi$PT calculation of Refs.~\cite{Gisbert:2017vvj,Cirigliano:2019cpi}, with a large-$N_C$ estimate of $a_{88}^{\lambda\lambda}$, and with the most recent lattice results~\cite{Abbott:2020hxn}.

Combining our analytical evaluation of the $K\to\pi\pi$ matrix elements~\cite{Gisbert:2017vvj,Cirigliano:2019cpi}, at NLO in $\chi$PT, with the numerical analysis of the RBC-UKQCD collaboration~\cite{Abbott:2020hxn},  we have extracted the leading chiral couplings through a direct fit to the lattice data. The comparison of these results, shown in Table~\ref{tab:fit}, with the corresponding large-$N_C$ estimates provides an enlightening anatomy of the well-known enhancement of the isoscalar $K\to\pi\pi$ amplitude, which we have discussed in detail in Section~\ref{subsec:lattice}. A dynamical QCD understanding of the so-called $\Delta I=\frac{1}{2}$ rule clearly emerges from this exercise.

The comparison with the lattice results also confirms that the $K\to\pi\pi$ matrix elements of the penguin operators $Q_6$ and $Q_8$
are well approximated by the large-$N_C$ limit, once the large $\chi$PT loop corrections (subleading in $1/N_C$) are properly taken into account. This was suggested long time ago \cite{Pallante:1999qf,Pallante:2001he}, based on the fact that the anomalous dimensions of these two operators are leading in $1/N_C$ and, moreover, the large-$N_C$ limit gives a good estimate of their exact values. The numerical confirmation of this property further reinforces the theoretical accuracy of the updated Standard Model prediction of $\varepsilon'/\varepsilon$ presented in Refs.~\cite{Gisbert:2017vvj,Cirigliano:2019cpi}, since $Q_6$ and $Q_8$ completely dominate the quantitative evaluation of this important observable.

Finally, we have also presented a beautiful consistency test between the experimental $\tau$-decay distribution, the $\chi$PT analytical description and the numerical lattice data. Using the lattice fit to determine the dimension-six condensate contribution to the $\Pi^d_{V-A}$ correlator, we have extracted the pion decay constant from the integrated $V-A$ invariant-mass distribution of the final hadrons in inclusive $\tau$ decays. The resulting value, given in Eq.~(\ref{eq:fpiTau}), is surprisingly accurate and in excellent agreement with the direct determinations from $\pi\to\mu\nu$ \cite{Zyla:2020zbs} and from lattice simulations \cite{Aoki:2019cca}.

\section*{Acknowledgements}

We are grateful to Hans Bijnens, Vincenzo Cirigliano, Hector Gisbert and Chris Sachrajda for useful discussions.
This work is partially supported by the Agence Nationale de la Recherche (ANR) under grant ANR-19- CE31-0012 (project MORA), by  the  Spanish  Government  and  ERDF funds from  the  European  Commission  (FPA2017-84445-P), by  the  Generalitat  Valenciana (PROMETEO/2017/053), by the EU H2020 research and innovation programme (Grant Agreement 824093) and by the EU COST Action CA16201 PARTICLEFACE.

\bibliographystyle{unsrtnat}
\bibliography{bib}

\end{document}